\PassOptionsToPackage{unicode}{hyperref}
\PassOptionsToPackage{hyphens}{url}
\PassOptionsToPackage{dvipsnames,svgnames,x11names}{xcolor}
\documentclass[
  12pt]{article}
\usepackage{wrapfig,lipsum,booktabs}
\usepackage{amsmath,amssymb}
\usepackage{iftex}
\ifPDFTeX
  \usepackage[T1]{fontenc}
  \usepackage[utf8]{inputenc}
  \usepackage{textcomp} 
\else 
  \usepackage{unicode-math}
  \defaultfontfeatures{Scale=MatchLowercase}
  \defaultfontfeatures[\rmfamily]{Ligatures=TeX, Scale=1}
\fi
\usepackage{lmodern}
\ifPDFTeX\else  
\fi
\IfFileExists{upquote.sty}{\usepackage{upquote}}{}
\IfFileExists{microtype.sty}{
  \usepackage[]{microtype}
  \UseMicrotypeSet[protrusion]{basicmath} 
}{}
\makeatletter
\@ifundefined{KOMAClassName}{
  \IfFileExists{parskip.sty}{%
    \usepackage{parskip}
  }{
    \setlength{\parindent}{0pt}
    \setlength{\parskip}{6pt plus 2pt minus 1pt}}
}{
  \KOMAoptions{parskip=half}}
\makeatother
\usepackage{xcolor}
\makeatletter
\ifx\paragraph\undefined\else
  \let\oldparagraph\paragraph
  \renewcommand{\paragraph}{
    \@ifstar
      \xxxParagraphStar
      \xxxParagraphNoStar
  }
  \newcommand{\xxxParagraphStar}[1]{\oldparagraph*{#1}\mbox{}}
  \newcommand{\xxxParagraphNoStar}[1]{\oldparagraph{#1}\mbox{}}
\fi
\ifx\subparagraph\undefined\else
  \let\oldsubparagraph\subparagraph
  \renewcommand{\subparagraph}{
    \@ifstar
      \xxxSubParagraphStar
      \xxxSubParagraphNoStar
  }
  \newcommand{\xxxSubParagraphStar}[1]{\oldsubparagraph*{#1}\mbox{}}
  \newcommand{\xxxSubParagraphNoStar}[1]{\oldsubparagraph{#1}\mbox{}}
\fi
\makeatother

\usepackage{longtable,booktabs,array}
\usepackage{calc} 
\usepackage{etoolbox}
\makeatletter
\patchcmd\longtable{\par}{\if@noskipsec\mbox{}\fi\par}{}{}
\makeatother
\IfFileExists{footnotehyper.sty}{\usepackage{footnotehyper}}{\usepackage{footnote}}
\makesavenoteenv{longtable}
\usepackage{graphicx}
\makeatletter
\def\maxwidth{\ifdim\Gin@nat@width>\linewidth\linewidth\else\Gin@nat@width\fi}
\def\maxheight{\ifdim\Gin@nat@height>\textheight\textheight\else\Gin@nat@height\fi}
\makeatother
\setkeys{Gin}{width=\maxwidth,height=\maxheight,keepaspectratio}
\makeatletter
\def\fps@figure{htbp}
\makeatother

\makeatletter
\@ifpackageloaded{caption}{}{\usepackage{caption}}
\AtBeginDocument{%
\ifdefined\contentsname
  \renewcommand*\contentsname{Table of contents}
\else
  \newcommand\contentsname{Table of contents}
\fi
\ifdefined\listfigurename
  \renewcommand*\listfigurename{List of Figures}
\else
  \newcommand\listfigurename{List of Figures}
\fi
\ifdefined\listtablename
  \renewcommand*\listtablename{List of Tables}
\else
  \newcommand\listtablename{List of Tables}
\fi
\ifdefined\figurename
  \renewcommand*\figurename{Figure}
\else
  \newcommand\figurename{Figure}
\fi
\ifdefined\tablename
  \renewcommand*\tablename{Table}
\else
  \newcommand\tablename{Table}
\fi
}
\@ifpackageloaded{float}{}{\usepackage{float}}
\floatstyle{ruled}
\@ifundefined{c@chapter}{\newfloat{codelisting}{h}{lop}}{\newfloat{codelisting}{h}{lop}[chapter]}
\floatname{codelisting}{Listing}

\makeatother
\makeatletter
\@ifpackageloaded{caption}{}{\usepackage{caption}}
\@ifpackageloaded{subcaption}{}{\usepackage{subcaption}}
\makeatother

\ifLuaTeX
  \usepackage{selnolig}  
\fi
\usepackage[]{natbib}
\usepackage{bookmark}

\IfFileExists{xurl.sty}{\usepackage{xurl}}{} 
\hypersetup{
  pdftitle={Title},
  pdfauthor={Author 1; Author 2},
  pdfkeywords={3 to 6 keywords, that do not appear in the title},
  colorlinks=true,
  linkcolor={blue},
  filecolor={Maroon},
  citecolor={Blue},
  urlcolor={Blue},
  pdfcreator={LaTeX via pandoc}}

\usepackage{amssymb,amsthm,amsmath}
\usepackage{xcolor,paralist,hyperref,fancyhdr,etoolbox}
\usepackage{xr}
\usepackage{algorithm, algorithmic}
\usepackage{siunitx}
\usepackage{svg} 
\usepackage{makecell}
\usepackage{multirow}
\usepackage{booktabs}
\usepackage{amsfonts}
\usepackage{enumerate}
\usepackage{enumitem}
\usepackage{threeparttable}
\usepackage{natbib}
\usepackage{cases}
\usepackage{booktabs}
\usepackage{float}

\numberwithin{equation}{section}
\newtheorem{thm}{Theorem}
\newtheorem{cor}{Corollary}

\newtheorem{prop}{Proposition}
\newtheorem{ass}{\textbf{Assumption}}
\theoremstyle{definition}

\theoremstyle{remark}
\newtheorem{rem}{Remark}

\newtheorem{example}{Example}
\allowdisplaybreaks[2]

\newcommand{\Pm}{\mathbb{P}}

\newcommand{\anon}{1}

\begin{document}

\def\spacingset#1{\renewcommand{\baselinestretch}%
{#1}\small\normalsize} \spacingset{1}


\if1\anon
{
  \title{\bf Factorized Tail Volatility Model: Augmenting Excess-over-Threshold Method for High-Dimensional Hevay-Tailed Data}
  \author{Yifan Hu \\
    School of Data Science, Fudan University, Shanghai, China, 200433, \\
    and \\
    Yanxi Hou\thanks{
      Corresponding author. The authors gratefully acknowledge {that the work is supported by the National Natural Science Foundation of China Grants 72171055 and 71991471.}}\hspace{.2cm}\\ 
    School of Data Science, Fudan University, Shanghai, China, 200433. \\
    }
  \maketitle
} \fi

\if0\anon
{
  \bigskip
  \bigskip
  \bigskip
  \begin{center}
    {\LARGE\bf Factorized Tail Volatility Model: Augmenting Excess-over-Threshold Method for High-Dimensional Heavy-Tailed Data}
\end{center}
  \medskip
} \fi

\bigskip
\begin{abstract}
 Ecess-over-Threshold method is a crucial technique in extreme value analysis, which approximately models larger observations over a threshold using a Generalized Pareto Distribution. This paper presents a comprehensive framework for analyzing tail risk in high-dimensional data  by introducing the Factorized Tail Volatility Model (FTVM) and integrating it with central quantile models through the EoT method. This integrated framework is termed the FTVM-EoT method. In this framework, a quantile-related high-dimensional data model is employed to select an appropriate threshold at the central quantile for the EoT method, while the FTVM  captures heteroscedastic tail volatility by decomposing tail quantiles into a low-rank linear factor structure and a heavy-tailed idiosyncratic component. The FTVM-EoT method is highly flexible, allowing for the joint modeling of central, intermediate, and extreme quantiles of high-dimensional data, thereby providing a holistic approach to tail risk analysis. In addition, we develop an iterative estimation algorithm for the FTVM-EoT method and establish the asymptotic properties of the estimators for latent factors, loadings, intermediate quantiles, and extreme quantiles. A validation procedure is introduced, and an information criterion is proposed for optimal factor selection.  
Simulation studies demonstrate that the FTVM-EoT method consistently outperforms existing methods at intermediate and extreme quantiles.
\end{abstract}

\noindent%
{\it Keywords:} extreme value analysis; excess-over-threshold; factor model; heavy-tailed data
\vfill

\newpage
\spacingset{1.8} 

\section{Introduction}\label{sec:Introduction}

Tail risk analysis for high-dimensional data has emerged as an intriguing and significant area of research, where Extreme Value Theory (EVT) plays a central role in the development of both theoretical frameworks and inferential methodologies. Some recent studies have predominantly concentrated on the tail dependence of high-dimensional extremes and on dimension reduction techniques that effectively capture the intrinsic characteristics of the tails in high-dimensional datasets. For instance, \cite{nicolas2021detection} introduced the concept of sparse regular variation for high-dimensional extremes, which has shown great promise in capturing the dependence structure of extreme events. Building on this work, \cite{nicolas2020multivariate} further proposed the MUSCLE algorithm for clustering extremes, which provides a powerful tool for identifying tail dependence in high-dimensional data. \cite{chautru2015} proposed a dimension reduction technique for multivariate extreme value analysis, simplifying the analysis of complex high-dimensional data. \cite{cooley2019} introduced two decompositions for high-dimensional tail dependence using a transformed-linear algebra framework, developing a matrix of pairwise tail dependence metrics. \cite{drees2021} presented a principal component analysis (PCA) for multivariate extremes, effectively capturing key components while reducing dimensionality. Overall, these contributions have collectively advanced the field of extreme value analysis, offering enhanced tools and techniques for understanding and managing extreme events in high-dimensional data.

This paper focuses on the estimation of extreme quantiles of high-dimensional heavy-tailed data, which is another important research direction in extreme value statistics. It is well known that the \textit{Excess-over-Threshold} (or \textit{Peak-over-Threshold}) method is one classical estimation approach for univariate distributions in EVT. This method posits that the larger observations over a threshold approximately follow a Generalized Pareto Distribution (GPD). More specifically, let \( X \) be a random variable with distribution function \( F \), and let \( u \) denote a threshold. Then, the excess distribution \( F_u(x) := \mathbb{P}(X - u \leq x \mid X > u) \) satisfies the following relationship:
\begin{equation}\label{eq:pot}
   \lim_{u \uparrow x^*} \sup_{0\le x < x^*-u}|F_u(x)- G_{\gamma,\sigma}(x)| = 0, \quad\text{with}\quad   G_{\gamma,\sigma}(x):=1 - \left(1 + \gamma \frac{x}{\sigma}\right)_+^{-1/\gamma},
\end{equation}
where \( x^* \) is the right endpoint of \( F \), and \( G_{\gamma,\sigma}(x) \) is the GPD with two parameters: a scale parameter \( \sigma > 0 \) and a shape parameter \( \gamma \). The shape parameter \( \gamma \) is also referred to as the \textit{extreme value index} of \( F \). Based on \eqref{eq:pot}, both parametric and non-parametric estimation methods can be established, such as the maximum likelihood estimator and the moment estimator; see Section 3 of \cite{haan2006extreme}. The GPD limit in \eqref{eq:pot} alternatively states that the excess variable \( Y = X - u \) given \( X > u \) satisfies a decomposition between volatility and tail-heaviness component such that \( Y = \sigma \varepsilon \), where \( \varepsilon \) approximately follows \( G_{\gamma,1} \) 
Our study is driven by this motivation, with an extension to high-dimensional heavy-tailed data. Specifically, suppose \(\{Y_{i,t}\}_{1 \leq i \leq N, 1 \leq t \leq T}\) is a high-dimensional dataset satisfying \(Y_{i,t} = \sigma_{i,t}\varepsilon_{i,t}\), where \( \sigma_{i,t} \) represents the volatility component and \( \varepsilon_{i,t} \) represents the tail-heaviness component. However, to generalize the EoT approach for high-dimensional data, it is necessary to handle idiosyncratic effects of the model, namely $\sigma_{i,t}$ and $\varepsilon_{i,t}$, in high-dimensional extremes. This issue has not been addressed in the literature.. Consequently, developing a unified framework for predicting extreme risks for all $Y_{i,t}$, such as the extreme quantiles of the underlying distributions of $Y_{i,t}$, remains a significant challenge and an area of great interest. 

To address the idiosyncratic effects and propose inference methods for high-dimensional extremes, this paper introduces the \textit{Factorized Tail Volatility Model} (FTVM),
\begin{equation}\label{eq:ftvm}
 Y_{i,t} = l_{0i}^{\top} f_{0t} \varepsilon_{i,t},\quad 1\le i\le N,\,1\le t\le T.
\end{equation}
Here, $\sigma_{i,t}=l^\top_{0i}f_{0t}$ is a linear factor model with the factors \(\left\{f_{0t} \in \mathbb{R}^r, 1 \leq t \leq T\right\}\) and the loadings \(\left\{l_{0i} \in \mathbb{R}^r, 1 \leq i \leq N\right\}\). 
$\varepsilon_{i,t}$ are independent for \(1 \leq i \leq N\) and \(1 \leq t \leq T\), given the factors and loadings, and is generated from the distributions $\mathbb F_{i,t}$.  
For convenience, we denote the tail quantile functions \(U_{i,t}\) of \(\mathbb{F}_{i,t}\) as
\begin{equation}
 U_{i,t}(x) = \inf\left\{s \left| 1 - \mathbb{F}_{i,t}(s) \leq x^{-1} \right. \right\}, \quad x>0.
\end{equation}
and denote $\varepsilon_{i,t}=U_{i,t}(V_{i,t}^{-1})$, where \(V_{i,t}\) are independent and identically distributed (i.i.d.) uniform random variables on the interval \([0,1]\) for \(1 \leq i \leq N\) and \(1 \leq t \leq T\), given the factors and  loadings.
Thus, the \((1-\tau)\)-th quantile of $\mathbb F_{i,t}$ is $U_{i,t}(\tau^{-1})$. { In the FTVM, it is not necessary to assume that \( Y_{i,t} \) is positive. This assumption is not required because our objective is to analyze the tail risk of \( Y_{i,t} \), which represents the excess variable in high-dimensional data within the FTVM-EoT framework. Consequently, we assume that the tail quantile functions \( U_{i,t} \) to be tail-equivalent to a reference function \(U\), such that for each given \(N\) and \(T\),}
\[
 \sup_{1 \leq i \leq N} \sup_{1 \leq t \leq T} \left| \frac{U_{i,t}(x)}{U(x)} - 1 \right| \to 0 \quad \text{as } x \to \infty.
\]
This tail-equivalent condition specifies the tail-heaviness of the high-dimensional data $\{Y_{i,t}\}$. Thus, the FTVM possesses desirable properties by incorporating both heteroscedastic volatilities $\sigma_{i,t}$ and heteroscedastic extremes $\varepsilon_{i,t}$. Specifically, it specifies a linear factor model to capture the volatilities across all \(i\) and \(t\), and includes a tail-equivalent component \(U_{i,t}(V_{i,t}^{-1})\), which accounts for idiosyncratic  effects not explained by the factor model.

Our model is mainly related to two streams of literature. The first stream focuses on volatility modelling by factor models. For instance, \cite{barigozzi2020generalized} proposed a two-stage generalized dynamic factor model to analyze and forecast high-dimensional panels of economic time series, with a particular emphasis on both levels and volatilities. Similarly, \cite{DING2025105959} introduced a multiplicative volatility factor model to study the daily volatilities of a large number of stocks. This model effectively captures the co-movement of volatilities by incorporating a multiplicative common factor and an idiosyncratic variance exposure. The second stream of research addresses heteroscedastic extremes in extreme value analysis. \cite{Einmahl2016} expanded classical extreme value theory to accommodate non-identically distributed observations, specifically targeting heteroscedastic extremes where distribution tails vary proportionally. Their method, validated through simulations and real data analyses, highlights the significant impact of heteroscedasticity on extreme events. Building on this work, \cite{bucher2024statistics} extended the concept of heteroscedastic extremes to handle serially dependent observations, providing a local limit theorem for a kernel estimator of the scedasis function and a functional limit theorem for an estimator of the integrated scedasis function. Additionally, \cite{Hou2024} developed a two-stage method to predict extreme conditional quantiles in panel data, leveraging second-order conditions for heteroscedastic extremes. Their approach involves constructing a panel quantile regression model at an intermediate level and then extrapolating to an extreme level using extreme value theory. In summary, these two streams of research provide a solid foundation for our model by offering advanced methodologies to handle complex data structures and dependencies.

In our theoretical analysis of the FTVM, we classify (tail) quantile levels into three distinct categories: central (or fixed) quantile levels, intermediate quantile levels, and extreme quantile levels. Specifically, we define the intermediate (tail) quantile level as \(k/NT\), where \(k := k(NT) \to \infty\) and \(k / NT \to 0\) as both \(N \to \infty\) and \(T \to \infty\). For the quantile level \( p_{N,T} \) satisfying \( p_{N,T} = o(k / NT) \) as \(N \to \infty\) and \(T \to \infty\), we refer to \( p_{N,T} \) as the extreme (tail) quantile level. This paper primarily investigates the asymptotic properties of the intermediate and extreme (tail) quantiles of \(Y_{i,t}\), rather than the central quantiles, such as the 25\% or 50\% quantiles. Intermediate and extreme quantile levels serve distinct purposes and require different estimation approaches. Intermediate quantile levels, such as the 90\% or 95\% quantiles in practice, capture the behavior of relatively rare but not exceedingly uncommon events. In contrast, extreme quantile levels, such as the 99\% or 99.9\% quantiles, focus on the most exceptional and rare occurrences, which are crucial for assessing extreme tail risks.
Our study makes two key contributions. First, we develop an inference method for estimating intermediate and extreme quantiles in high-dimensional data and establish the asymptotic properties of the FTVM. The proposed model generalizes the classical EoT method in extreme value analysis and improves tail risk analysis for high-dimensional data by combining heteroscedastic volatilities and heteroscedastic extremes. Second, we integrate the FTVM with other popular high-dimensional quantile-related models, such as the Quantile Factor Model (QFM) proposed by \cite{Chen2021} and the Quantile Regression with Interactive Fixed Effects (QRIFE) introduced by \cite{ando2020quantile}. This integration aims to enhance tail risk analysis for existing high-dimensional models. Models like QFM and QRIFE, which focus on central quantiles, serve as a threshold model for high-dimensional data in our FTVM-EoT approach, while the FTVM is then applied to model the excess of central quantiles over these thresholds. Based on this approach, we can develop the asymptotic properties of multiple-stage inference methods, including extrapolation methods for extreme quantiles. Simulation studies demonstrate that models enhanced with the FTVM outperform their counterparts without the FTVM in terms of extreme risk analysis. While we demonstrate this approach using QFM and QRIFE in this paper, it holds promise for augmenting tail risk analysis in other high-dimensional quantile-related models.

The remainder of this article is structured as follows. Section \ref{sec:EQHP} outlines the detailed assumptions for the FTVM. Section \ref{sec:estimation} introduces methods for estimating the factors and loadings in the FTVM, including an iterative algorithm for solving the optimization problem and deriving the asymptotic properties of the optimized solution. Section \ref{subsec:MSM} presents a model validation method based on hypothesis testing using the Kolmogorov-Smirnov(KS) statistic. Additionally, we propose an estimator for determining the optimal number of factors using information criteria. Finally, Section \ref{sec:Location-scale-shift} introduces the Excess-over-threshold model, which combines the FTVM with other statistical tools to model the relationship between different quantile levels.

\subsection{Notations}
We define the notations used throughout the paper. 
We denote \(a \vee b = \max(a, b)\) for \(a, b \in \mathbb{R}\). The largest integer smaller than a real number \(a\) is denoted by \(\lfloor a \rfloor\). Let \(L_{0N} = (l_{01}, l_{02}, \ldots, l_{0N})\) and \(F_{0T} = (f_{01}, f_{02}, \ldots, f_{0T})\). Similarly, let \(L_{N,r} = (l_{1,r}, \ldots, l_{N,r})\) denote a matrix in \(\mathbb{R}^{r \times N}\), and \(F_{T,r} = (f_{1,r}, \ldots, f_{T,r})\) denote a matrix in \(\mathbb{R}^{r \times T}\). 
Let \(\mathbb{I}_r\) denote the \(r \times r\) identity matrix, and \(\operatorname{diag}(a_1, a_2, \ldots, a_r)\) denote the \(r \times r\) diagonal matrix with diagonal elements \(a_1, a_2, \ldots, a_r\). Let \(\mathbf{1}^{N}\) denote a \(1 \times N\) vector with all elements equal to 1. For a real number \(a\), define \(\operatorname{sgn}(a) = 1\) if \(a \geq 0\) and \(\operatorname{sgn}(a) = -1\) if \(a < 0\). For a matrix \(A \in \mathbb{R}^{r \times r}\), \(\operatorname{sgn}(A)\) is defined as the diagonal matrix whose \(j\)-th diagonal element equals the \(\operatorname{sgn}\) of the \(j\)-th diagonal element of \(A\). 
The infinity norm of a matrix is denoted by \(\Vert \cdot \Vert_\infty\), and the Frobenius norm is denoted by \(\Vert \cdot \Vert_{\operatorname{F}}\).  In our paper, we analyze the weak convergence of \(Z_{N, T}\) to \(Z\) as \(N \to \infty\) and $T\to \infty$ given $L_{0N}$ and $F_{0T}$, which is denoted by \(Z_{N, T} \leadsto Z\).

\section{Factorized Tail Volatility Model}
\label{sec:EQHP}

In this section, we present the detailed assumptions for the FTVM. We first need an identification assummption on the volatility componet of the FTVM.

\begin{ass}[Identification Constraints]\label{ass: identify} 
 For $N$, $T > 0$ and all $1 \le i \le N$, $1 \le t\le T$,  there exist compact sets $\mathcal{L}$ and $\mathcal{F}$ such that  $ l_{0i} \in \mathcal{L}$ and $f_{0t} \in \mathcal{F}$. 
 There exists positive constants $m$, $M$ such that \( m \le 1 \le M \) and  for all \( N, T > 0 \),
    \[
 m \leq
 \inf_{\substack{1 \leq i \leq N, \, 1 \leq t \leq T}} l_{0i}^\top f_{0t}  
        \leq 
 \sup_{\substack{1 \leq i \leq N, \, 1 \leq t \leq T}} l_{0i}^\top f_{0t}  
        \leq M.
    \]
 
 The loading matrix satisfies that as \( N \to \infty \),
    \[
 N^{-1} L_{0N} L_{0N}^{\top}
 = \operatorname{diag}\left(
        \sigma_{N 1}, \ldots, \sigma_{N r}
 \right)
 \to \operatorname{diag} (\sigma_1, \sigma_2, \ldots, \sigma_r),
    \]
 where \( \sigma_{N 1} \geq \sigma_{N 2} \geq \ldots \geq \sigma_{N r} \)
 and \( \infty > \sigma_1 > \sigma_2 > \ldots > \sigma_r > 0 \).

 The factor matrix satisfies that for each \( T > 0 \), 
    $
 T^{-1} F_{0T} F_{0T}^{\top} = \mathbb{I}_r.
   $
\end{ass}

Assumption \ref{ass: identify} is similar to the identification assumption in \cite{ando2020quantile} and \cite{Chen2021}.
Additionally, we assume that \( l_{0i}^\top f_{0t} \) is bounded away from zero, which is necessary to ensure that all quantiles of \( Y_{i,t} \) are non-degenerate functions.

\begin{ass}[Heteroscedastic Tail Quantile]\label{ass:HEQ}
 Suppose the distribution functions \( \mathbb{F}_{i,t} \) are continuous.
 The functions \( U_{i,t} \) are tail-equivalent to a reference function \( U \)
 such that for a series of positive and decreasing function \( A_{N,T} \), 
    \begin{equation}
 \sup_{N, T \in \mathbb{N}}
 \sup_{\substack{1 \leq i \leq N, \, 1 \leq t \leq T}} \sup_{x > NT/k(2M^{1/\gamma})}
 \left|\frac{U_{i,t}(x) / U(x) - 1}{A_{N,T}(x)}\right| \le C_0.
    \end{equation}
 The reference function \( U \) has an extreme value index \( \gamma > 0 \) such that for all \( x > 0 \), a \( \rho < 0 \), and an eventually decreasing function \( A_1 \),
    \begin{equation}\label{eq: ht}
 \lim_{s \to \infty }
 \frac{1}{A_1(s)}
 \left(
 \frac{U(sx)}{U(s)} - x^\gamma
 \right)
 = x^{\gamma} \frac{x^\rho - 1}{\rho}.
    \end{equation}

 Moreover, as \( N\to \infty \) and $T\to \infty$,
    \begin{equation}
        \label{eq: mean}
 \sqrt{k} \left|\frac{1}{NT} \sum_{i = 1}^{N} \sum_{t=1}^{T}
 (l_{i0}^\top f_{t0})^{1/\gamma} - 1\right| \to 0.
    \end{equation}
\end{ass}

\begin{ass}[Intermediate Order]\label{ass:intermediate} 
 The sequence \( k = k(NT) \) satisfies \( k/NT \to 0 \),
    \( (N + T)/k \to 0 \), \( \sqrt{k} A_{N,T}(NT m^{1/\gamma}/(4\, 2^{1/\gamma}  M^{2/\gamma} k)) \to 0 \), and 
    \( \sqrt{k} A_1(NT/k) \to 0 \) as \( N \to \infty \) and \( T \to \infty \).
\end{ass}

Assumptions \ref{ass:HEQ} and \ref{ass:intermediate} represent special cases of heterogeneous extremes as discussed in \cite{Einmahl2023}. 
Additionally, the constraint \eqref{eq: mean} is consistent with the framework used in \cite{Einmahl2016}. 
These assumptions enable the estimation of the reference quantile \( U(NT/k) \) without explicit knowledge of the factorized volatility structure, as demonstrated in the following proposition. We denote the Hill estimator as 
\[
\hat{\gamma} :=
k^{-1} \sum_{i=1}^{k} \log\{\hat{U}(NT/i)\} - \log\{\hat{U}(NT/k)\},
\]
where  \(\hat{U}(NT/k)\) is denoted as the \(k\)-th largest order statistic of \(\{Y_{i,t}\}_{1 \leq i \leq N, 1 \leq t \leq T}\).
\begin{prop}\label{prop:qest}
 Under Assumptions \ref{ass:HEQ} and \ref{ass:intermediate}, as \( N \to \infty \) and \( T \to \infty \),
 \begin{enumerate}
   \item for the $k$-th largest order statistic \( \hat{U}(NT/k) \), it holds that
   \[
\sqrt{k} \left(\frac{\hat{U}(NT/k)}{U(NT/k)} - 1\right)
       \leadsto N(0, \gamma^2).
   \]
   \item for the Hill estimator $\hat{\gamma}$, it holds that
   $
\sqrt{k} \left(\hat{\gamma} - \gamma\right) \leadsto N(0, \gamma^2).
   $
 \end{enumerate}

\end{prop}

In practical applications, the reference \( U(NT/k) \) can be interpreted as the unconditional tail quantile of the sequence \(\{Y_{i,t}\}\). To elaborate, consider a scenario where \( l_{0i} \) and \( f_{0t} \) are i.i.d. latent random vectors. Assume that \( V_{i,t} \) is independent of \( l_{0i} \) and \( f_{0t} \), and that \( U_{i,t} = U \) for all \( 1 \leq i \leq N \) and \( 1 \leq t \leq T \). Let \( \mathbb{F} \) denote the cumulative distribution function of \( U \). For the random variable \( Y_{i,t} = l_{0i}^\top f_{0t} U(V_{i,t}^{-1}) \), the following holds:
\begin{align*}
 \mathbb{P}(Y_{i,t} > U(NT/k))
 = & \mathbb{E} \left[\mathbb{P}(l_{0i}^\top f_{0t} U(V_{i,t}^{-1}) > U(NT/k) \mid l_i, f_t)\right]  \\
 = & \left\{1 - \mathbb{F}(U(NT/k))\right\} \mathbb{E} \left[ \frac{1 - \mathbb{F}((l_{0i}^\top f_{0t})^{-1}U(NT/k))}{1 - \mathbb{F}(U(NT/k))} \right]  \\
    \approx & \frac{k}{NT} \mathbb{E} \left(l_{0i}^\top f_{0t}\right)^{1/\gamma} \approx \frac{k}{NT}.
\end{align*}
The penultimate approximation is derived from \eqref{eq: ht} and Theorem 2.3.9 in \citet{haan2006extreme}, while the final step follows from \eqref{eq: mean} and the assumption that \( l_{0i} \) and \( f_{0t} \) are i.i.d.. Consequently, for sufficiently large \( N \) and \( T \), \( U(NT/k) \) asymptotically represents the unconditional tail quantile in the FTVM.


Before studying the asymptotic convergence, we first discuss several closely related models.

\begin{example}[Location-Scale-Shift Model]\label{ex:LSSM}
 \cite{Chen2021} proposes a special case of quantile factor model of the form:
    \[
 Y_{i,t} = \alpha_{0i}^\top \beta_{0t} + l_{0i}^\top f_{0t} U(V_{i,t}^{-1}).
    \]
 When \( \alpha_{0i}^\top \beta_{0t} \) is bounded for \( 1 \le i \le N \) and \( 1 \le t \le T \), the tail quantile of \( Y_{i,t} \) satisfies:
    \[
 \lim_{NT \to \infty} \max_{\substack{1 \le i \le N \\ 1 \le t \le T}}
 \left| \frac{\alpha_{0i}^\top \beta_{0t} + l_{0i}^\top f_{0t} U(NT/k)}{U(NT/k)} - l_{0i}^\top f_{0t} \right| = 0.
    \]
 This implies that intermediate tail quantiles are asymptotically dominated by the heterogeneous term \( l_{0i}^\top f_{0t} U(V_{i,t}^{-1}) \), consistent with the structure of FTVM. 

 For an extreme tail quantile level, it holds that
    \[
 \lim_{NT \to \infty} \max_{\substack{1 \le i \le N \\ 1 \le t \le T}}
 \left| \frac{\alpha_{0i}^\top \beta_{0t} + l_{0i}^\top f_{0t} U(NT/k) \left(k / (NTp_{N,T})\right)^{\gamma}}{ U((p_{N,T})^{-1})} - l_{0i}^\top f_{0t} \right| = 0.
    \]
 Thus, the extreme tail quantile is asymptotically equivalent to \( l_{0i}^\top f_{0t} U(NT/k) \left(k / (NTp_{N,T})\right)^{\gamma} \).

 Although we assume \( N \to \infty \) and \( T \to \infty \), in practical applications, the scale of the intermediate tail quantile \( l_{0i}^\top f_{0t} U(NT/k) \) may be comparable to the bounded term \( \alpha_{0i}^\top \beta_{0t} \) when \( NT / k \) is not sufficiently large. For example, consider the intermediate tail quantile level at 5\% with \( N = 100 \), \( T = 100 \), and \( k = 500 \). For a \( t \)-distribution with degree of freedom  3, the 5\% tail quantile is approximately 2.35. If \( \alpha_{0i}^\top \beta_{0t} \) is around 3, it becomes difficult to distinguish between \( \alpha_{0i}^\top \beta_{0t} \) and \( l_{0i}^\top f_{0t} U(V_{i,t}^{-1}) \). 
 Furthermore, if we mistakenly apply the extrapolation,
    \[
 \left(\alpha_{0i}^\top \beta_{0t} + l_{0i}^\top f_{0t} U(NT/k)\right) \left(k / (NTp_{N,T})\right)^{\gamma},
    \]
 the estimated extreme tail quantile becomes unreliable. We will further explore this challenge as an application of FTVM in Section \ref{sec:Location-scale-shift}.
\end{example}

\begin{example}[Two-Way Fixed Effect Model]
 A common approach for data transformation is to apply the Box-Cox transformation and analyze the statistical properties of the transformed variables. For the FTVM with a single factor, the Box-Cox transformation with a parameter of \(0\) results in the following decomposition:
\[
\log(Y_{i,t}) = \log(l_{0i}) + \log(f_{0t}) + \log(U(V^{-1}_{i,t})),
\]
which separates the logarithm of the tail quantile into an individual fixed effect \(\log(l_{0i})\), a time fixed effect \(\log(f_{0t})\), and a common term \(\log(U(V^{-1}_{i,t}))\). 
It is important to note that the transformed varaible \(\log(Y_{i,t})\) has an extreme value index zero, which makes the estimation of extreme and intermediate tail quantiles more challenging. This difficulty arises because additional constants must be estimated to derive the asymptotic results (see, for example, Lemma 3.5.5 and Theorem 4.3.1 in \citet{haan2006extreme}). However, as demonstrated in Proposition \ref{prop:qest} and Theorem \ref{thm:asymrate}, the estimation of extreme and intermediate tail quantiles under the FTVM framework is more straightforward. Therefore, we recommend applying the FTVM to identify the factors and loadings, as it simplifies the estimation process.
\end{example}



\section{Estimators of Factors and Loadings}\label{sec:estimation}

Suppose the number of factors \( r \) is known. We estimate \( L_{0N} \) and \( F_{0T} \) by solving the following optimization problem:
\begin{align}
 (\hat{L}_{N,r}, \hat{F}_{T,r})  = &
 (\hat{l}_{1,r}, \ldots, \hat{l}_{N,r}, \hat{f}_{1,r}, \ldots, \hat{f}_{T,r})\label{eq:problem2}
    \\
 =                                 & \arg\min_{{L_{N,r}}, {F_{T,r}}} 
  \sum_{i=1}^{N} \sum_{t=1}^{T}
    \rho_{(k/NT)} \left(
 \frac{Y_{i,t}}{\hat{U}(NT/k)} -{l}_{i,r}^{\top}{f}_{t,r}
 \right),  \nonumber                                                                     \\
& \begin{array}{cll}
\text{s.t. }                                &
m < {l}_{i,r}^{\top}{f}_{t,r} \le M,   &
\text{for \( i =1,\ldots, N \), and \( t = 1,\ldots, T \)}.
\end{array} \nonumber
\end{align}
Here, \(\rho_{(k/NT)}\) is the check function defined as
$
\rho_{(\tau)}(x) := ( \mathbf{1}(x > 0)- \tau) x,
$
which is used to minimize the loss at the $\tau$-th tail quantile. 
We propose an iterative algorithm to solve \eqref{eq:problem2} in Algorithm \ref{alg:TQFM2} in the supplementary material. 
We then derive the asymptotic properties of the optimized solution to \eqref{eq:problem2}.
To proceed, we define the following \textit{Mean Squared Relative Error }(MSRE) for \( L_{N,r} \), \( F_{T,r} \), \(\Lambda\) and a tail quantile level $\tau$, 
\begin{align}\label{eq:MSREdef}
 \operatorname{MSRE}_{\tau} (L_{N,r}, F_{T,r}, \Lambda)
     & :=
 {\frac{1}{NT}
 \sum_{i=1}^{N} \sum_{t=1}^{T} \left(
{ \frac{{l}_{i,r}^\top {f}_{t,r} \Lambda}{U(\tau^{-1})}
} - \frac{{l}_{0i}^\top {f}_{0t}{{U}_{i,t}(\tau^{-1})}}{{U(\tau^{-1})}}
\right)^2}. 
\end{align}



\begin{thm}\label{thm:asymrate}
 Under Assumptions \ref{ass: identify}-\ref{ass:intermediate}, 
 suppose $p_{N,T}$ is an extreme tail quantile level such that \( NTp_{N,T} = o(k) \) and \( \log(NTp_{N,T}) = o(\sqrt{k}) \) as \( N \to \infty \) and \( T \to \infty \). Then, 
 \begin{enumerate}
   \item for the estimators of loadings and factors, 
 it holds that for \( \hat{S} = \operatorname{sgn}{F}_{0T}\hat{F}_{T,r}^\top \),
       \[
  N^{-1/2} \|\hat{L}_{N,r} - \hat{S} {L}_{0N}\|_{\operatorname{F}} = O_p\left(\sqrt{\frac{N+T}{k}} \right)\;\text{and}\; T^{-1/2} \|\hat{F}_{T,r} - \hat{S} {F}_{0T}\|_{\operatorname{F}}  = O_p\left(\sqrt{\frac{N+T}{k}} \right).
\]
          \item for the intermediate tail quantile factorization, it holds that
   \begin{align*}
 \operatorname{MSRE}_{k/NT} (\hat{L}_{N,r}, \hat{F}_{T,r}, \hat{U}(NT/k))
     & = O_p\left({{\frac{N+T}{k}}} \right).
   \end{align*}
   \item for the extreme tail quantile factorization, it holds that
   \begin{align*}
    \operatorname{MSRE}_{p_{N,T}} \left(\hat{L}_{N,r}, \hat{F}_{T,r}, \hat{U}(NT/k) \left(\frac{k}{NT p_{N,T}}\right)^{\hat{\gamma}} \right) = O_p\left({\frac{N+T}{k}} \vee \frac{\log^2(k/(NT p_{N,T}))}{{k}}\right).
  \end{align*}
 \end{enumerate}
    
\end{thm}

\begin{rem}\label{rem:1}
 Consider the case when \( N = T \) and an appropriate intermediate rate \( k \). It is important to note that in each iteration of the algorithm, the optimization problem involves fitting a quantile regression at the tail quantile level \( k/(NT) \). By \citet{chernozhukov2017extremal}, if the ground truth of \( l_{0i} \) is known, the best estimator for \( f_{0t^*} \) is obtained by conducting quantile regression on the data \( \{Y_{i,t}\}_{t = t^*, 1 \le i \le N} \) at the intermediate tail quantile level \( k / NT \), whose convergence rate is achieved as \( \sqrt{N/k} \). In this regard, the convergence rate of FTVM aligns with the results of \citet{chernozhukov2017extremal}.
\end{rem}

\begin{rem}
   
A key aspect of the optimization problem \eqref{eq:problem2} is that we bound the intermediate tail quantiles of each \( Y_{i,t} \) around the unconditional tail quantile of the entire data, \( \hat{U}(NT/k) \). This constraint is necessary for proving the consistency of the estimators. Specifically, in the proof, we apply Proposition \ref{prop:uineq} in the supplementary material repeatedly to bound
\[
(NT)\{1-\mathbb{F}_{i,t}\left(  {U}_{i,t}(NT/k) (1+s)\right) \}/k
-  (1+s)^{-1/\gamma} 
\]
for \( s \) related to \( \hat{l}_{i,r}^\top \hat{f}_{t,r}  \hat{U}(NT/k) / ({l}_{0i}^\top {f}_{0t} U(NT/k)) \). Since Proposition \ref{prop:uineq} is derived only for \( x \) in a compact set, the constraint of \eqref{eq:problem2} is thus necessary.

\end{rem}


\section{Model Validation and Factor Selection}\label{subsec:MSM}

In this section, we propose a method of model validation and factor selection for the FTVM. In Section \ref{sec:estimation}, we observe that the MSRE for the estimators \(\hat{L}_{N,r}\) and \(\hat{F}_{T,r}\) converges at a relatively slow rate. For instance, when \(N = 50\), \(T = 50\), and \(k = 125\), the ratio \((N+T)/k = 0.8\) is significantly larger than \(1/N = 0.02\). This indicates that while a \(50 \times 50\) data is sufficient for a good estimation of central tail quantiles, as demonstrated in the experiments of \cite{Chen2021}, the performance of FTVM at intermediate tail quantiles may be suboptimal. In such cases, the unconditional tail quantile estimator \(\hat{U}(NT/k)\) might outperform FTVM in estimating the tail quantiles of \(Y_{i,t}\).

To address this issue, we propose a systematic approach for validating the applicability of the FTVM and selecting the optimal number of factors. We begin by introducing the degenerate FTVM, 
\begin{equation}\label{eq:baseline}
 H_0: Y_{i,t} = U_{i,t}(V_{i,t}^{-1}),\quad\text{for all }i\text{ and }t,
\end{equation}
 A hypothesis test is then developed to determine whether this degenerate FTVM is suitable for the given data. If the test indicates a non-degenerate FTVM is appropriate, we further propose an information criterion-based method to estimate the optimal number of factors.

The degenerate FTVM \eqref{eq:baseline} represents a simplified version of the FTVM, where the factors and loadings remain constant across all observations. This simplification is particularly useful in scenarios where the data lacks strong heterogeneity. Here, ``strong heterogeneity" refers to cases where \( {(N+T)/k} \) is significantly smaller than \( {{NT}^{-1}\sum_{i=1}^{N} \sum_{t=1}^{T} \left( 1 - {l}_{0i}^\top {f}_{0t} \right)^2} \). Specifically, when \( N \) and \( T \) are small, heterogeneity may not be strong. In such cases, the MSRE of the degenerate FTVM can be calculated as:
\begin{align*}
    & \operatorname{MSRE_{k/NT}}(\mathbf{1}^{N}, \mathbf{1}^{T}, \hat{U}(NT/k)) \\
 = & \frac{1}{NT} \sum_{i=1}^{N} \sum_{t=1}^{T} \left(
 \frac{\hat{U}(NT/k) - {l}_{0i}^\top {f}_{0t} {U}_{i,t}(NT/k)}{U(NT/k)}
 \right)^2 \\
    \leq & \frac{2}{NT} \sum_{i=1}^{N} \sum_{t=1}^{T} \left(
 \frac{\hat{U}(NT/k)}{U(NT/k)} - \frac{{U}_{i,t}(NT/k)}{U(NT/k)}
 \right)^2 + \frac{2}{NT} \sum_{i=1}^{N} \sum_{t=1}^{T} \left(
 \frac{{U}_{i,t}(NT/k)}{U(NT/k)} \right)^2 \left( 1 - {l}_{0i}^\top {f}_{0t} \right)^2 \\
    \leq & O_p\left(\frac{1}{k}\right) + \frac{4}{NT} \sum_{i=1}^{N} \sum_{t=1}^{T} \left( 1 - {l}_{0i}^\top {f}_{0t} \right)^2 \\
    \lesssim & \operatorname{MSRE}_{k/NT}(\hat{L}_{N,r}, \hat{F}_{T,r}, \hat{U}(NT/k)).
\end{align*}
Thus, the hypothesis test between the degenerate FTVM and a standard FTVM can be interpreted as a test of whether the heterogeneity in the high-dimensional data is strong enough to justify the use of FTVM. If the heterogeneity is weak, the degenerate FTVM may provide a simpler and more effective model for estimating tail quantiles.


To this end, we propose the following KS statistic:
\begin{align*}
 \operatorname{KS} := & \sup_{0 \le s \le 1} \sqrt{k}\,\, \Bigg| 
 \left\{ \frac{1}{k} \sum_{t=1}^{\lfloor T s \rfloor} \sum_{i=1}^{N} 
 \mathbf{1}\left( Y_{i,t} \ge \hat{U}\left({NT}/{k}\right) \right) \right\} \\
   & +
 \left\{ \frac{1}{k} \sum_{i=1}^{\lfloor NTs \rfloor - N\lfloor T s \rfloor}
 \mathbf{1}\left( Y_{i,\lfloor Ts \rfloor + 1} \ge 
 \hat{U}\left({NT}/{k}\right) \right) 
 \right\} - s
 \Bigg|.
   \end{align*}

\begin{prop}\label{prop:test}
 Under Assumptions \ref{ass: identify}-\ref{ass:intermediate} and \( H_0 \), { there exists a standard Brownian Bridge \( B \) on $[0,1]$} such that as \( N \), \( T \to \infty \),
    $
 \operatorname{KS} \leadsto \sup_{ 0 \le s \le 1} |B(s)|.
    $
\end{prop}

If \( H_0 \) is rejected, we then provide the following estimator to determine the optimal number of factors for FTVM. We estimate
\[
\hat{L}_{N,l} = (\hat{l}_{1,l}, \ldots, \hat{l}_{N,l})
    \in \mathbb{R}^{l \times N} \quad \text{and} \quad
\hat{F}_{T,l} = (\hat{f}_{1,l}, \ldots, \hat{f}_{T,l})
    \in \mathbb{R}^{l \times T}
\]
by solving \eqref{eq:problem2} with \( \hat{l}_{i,l} \) and \( \hat{f}_{t,l} \) as \( l \)-dimensional vectors instead of \( r \)-dimensional ones.
The \textit{Information Criteria} is proposed for the estimation of factor number:
\begin{equation}\label{eq:ric}
 \hat{r}_{\mathrm{IC}} = \arg\min_{1 \le l \le r^*}
 \frac{1}{k} \sum_{i=1}^{N} \sum_{t=1}^T \rho_{(k/NT)}
 \left(\frac{{Y_{i,t}}}{\hat{U}(NT/k)} -
 \hat{l}_{l,i}^\top \hat{f}_{l,t}
 \right) + l \cdot P_{N,T},
\end{equation}
where \( P_{N,T} = P_{N,T}(N,T) \) is a specific threshold related to \( N,T \), and $r^*$ is a sufficient large constant satisfying $r < r^*$.

\begin{thm}\label{thm:selection}
 Under Assumptions \ref{ass: identify}-\ref{ass:intermediate}, suppose \(P_{N,T} (k /(N+T)) \to \infty\) and \(P_{N,T} \to 0\) as \(N, T \to \infty\). It holds that as \( N \to \infty \) and \( T \to \infty \), 
   $
 \Pm(\hat{r}_{\mathrm{IC}} = r) \to 1.
   $
\end{thm}

To summarize, we recommend the following steps for model validation and factor selection:
\begin{enumerate}
   \item Conduct the hypothesis test \( H_0 \) using the KS statistic.
   \item If \( H_0 \) is not rejected, apply \( \hat{U}(NT/k) \) and $\hat{U}(NT/k) ({k}/{(NT p_{N,T})})^{\hat{\gamma}}$ as the estimator of the intermediate and extreme tail quantiles of \( Y_{i,t} \).
   \item If \( H_0 \) is rejected, use \( \hat{r}_{\mathrm{IC}} \) to determine the optimal number of factors.
\end{enumerate}


\begin{rem}
If we optimize \(\hat{r}_{\mathrm{IC}}\) for \(0 \leq l \leq r^*\), and the estimated \(\hat{r}_{\mathrm{IC}}\) equals \(0\), the selection method defaults to the degenerate FTVM. 
Thus, the degenerate FTVM can be regarded as a special case of the FTVM with \(r=0\).
In the proof, we show that when the data \(Y_{i,t}\) is generated under \(H_0\), the probability \(\Pm(\hat{r}_{\mathrm{IC}} = 0) \to 1\) holds as \(N, T \to \infty\). However, we still recommend first conducting the hypothesis test and then selecting the appropriate number of factors. This is because the selection process, which involves fitting the FTVM with various factor numbers, is computationally expensive. Additionally, the hypothesis test provides strong explanatory power for determining the applicability of the FTVM and the heterogeneity of the high-dimensional data. As a result, the hypothesis test is more interpretable and efficient compared to the selection process.
\end{rem}

\begin{rem}
A more refined formulation involves choosing \(P_{N,T}\) as follows:
\begin{equation}\label{eq:PPNT}
P_{N,T} = \left(\frac{N+T}{c \, k}\right) \log{\left(\frac{k}{N+T}\right)} \left\{
\frac{1}{k} \sum_{i=1}^{N} \sum_{t=1}^{T}\rho_{(k/NT)}\left(\frac{Y_{i,t}}{\hat{U}(NT/k)} - 1\right)
\right\},
\end{equation}
where \(c\) is a chosen constant, and \(\left({(N+T)}/{k}\right) \log{\left({k}/{N+T}\right)}\) serves as the penalty term introduced in \cite{ando2020quantile}. The penalty term is scaled for the following reasons. 

First, $
{k}^{-1} \sum_{i=1}^{N}\sum_{t=1}^{T}\rho_{(k/NT)}\left({Y_{i,t}}/{\hat{U}(NT/k)} - 1\right) = O_p(1)$ as \(N, T \to \infty\) and is bounded away from zero, as shown in Proposition~\ref{prop:decompose} in the supplementary material. This ensures the conditions for Theorem~\ref{thm:selection} are satisfied. 
Second, empirical observations indicate that the scale of the loss function in optimization problem~\eqref{eq:problem2} varies for different settings. To address this, we set \(L_{N,1} = \mathbf{1}^N\) and \(F_{T,1} = \mathbf{1}^T\) in the loss function of~\eqref{eq:problem2} to determine the appropriate scaling. 
Finally, the information criterion \eqref{eq:ric} can be rewritten as:
\[
\hat{r}_{\mathrm{IC}} = \arg\min_{0 \le l \le r^*}
\frac{\sum_{i=1}^{N} \sum_{t=1}^T \rho_{(k/NT)}
\left(\frac{{Y_{i,t}}}{\hat{U}(NT/k)} -
\hat{l}_{l,i}^\top \hat{f}_{l,t}
\right)}{
\sum_{i=1}^{N} \sum_{t=1}^{T}\rho_{(k/NT)}\left(\frac{Y_{i,t}}{\hat{U}(NT/k)} - 1\right)
} + l \left(\frac{N+T}{c \, k}\right) \log{\left(\frac{k}{N+T}\right)}.
\]
The first term is analogous to the fraction of variance unexplained in linear regression, except that the check function is used instead of the squared loss function. The second term acts as the penalty term. The optimal number of factors is then determined by balancing the trade-off between the goodness-of-fit criterion and the penalty.
\end{rem}

\section{FTVM-EoT Approach}
\label{sec:Location-scale-shift}

In this section, we introduce the FTVM-EoT method, a general framework that integrates the FTVM with other statistical models, { denoted as \(\mathcal{H}_{0,\tau^*}\)}, to enhance the estimation of extreme tail quantiles. 
Specifically, we analyze the conditional \(\tau\)-th tail quantile of \(Y_{i,t}\), denoted as \( \tilde{U}_{i,t}\left({\tau}^{-1} \mid \mathcal{I}_{i,t}, f_{0t}, l_{0i} \right)\), given the conditioning variables \(\mathcal{I}_{i,t}\), \(f_{0t}\), and \(l_{0i}\).

In the FTVM-EoT framework, \(\mathcal{H}_{0,\tau^*}\), is referred to as the threshold model for a central tail quantile level \(\tau^*\), and it is also a conditional quantile model based on the information set \(\mathcal{I}_{i,t}\) such that
\begin{equation}
   \tilde{U}_{i,t}((\tau^*)^{-1} \mid \mathcal{I}_{i,t}) := \mathcal{H}_{0,\tau^*}(\mathcal{I}_{i,t}),
\end{equation}
where \(\mathcal{I}_{i,t}\) may include explanatory variables or factors. The FTVM-EoT method encompasses a class of high-dimensional models by allowing flexibility in the choice of the threshold model. For example, if the Quantile Factor Model (QFM) is selected as the threshold model, the resulting implementation is referred to as QFM-FTVM, where the information set corresponds to the factors in the QFM. Similarly, if the Quantile Regression with Interactive Fixed Effects (QRIFE) is chosen, the implementation is denoted as QRIFE-FTVM, where the information set includes the explanatory variables in the QRIFE.

The quantiles exceeding the threshold model are then modeled using the FTVM. Specifically, the conditional quantiles \(\tilde{U}_{i,t}(\tau^{-1} \mid \mathcal{I}_{i,t}, f_{0t}, l_{0i})\) for \(\tau > \tau^*\) are assumed as
\begin{equation}
   \tilde{U}_{i,t}(\tau^{-1} \mid \mathcal{I}_{i,t}, f_{0t}, l_{0i}) 
:=   \tilde{U}_{i,t}((\tau^*)^{-1} \mid \mathcal{I}_{i,t})+\Delta \tilde{U}_{i,t}(\tau^{-1} \mid f_{0t}, l_{0i}),
\end{equation}
where the tail quantiles of \(\Delta \tilde{U}_{i,t}(\tau^{-1} \mid f_{0t}, l_{0i})\) are modeled using the FTVM,
\begin{equation}
   \Delta \tilde{U}_{i,t}(\tau^{-1} \mid f_{0t}, l_{0i}) := l_{0i}^\top f_{0t} {U}_{i,t}(\tau^{-1}).
\end{equation}
In the absence of specific assumptions, it is challenging to connect the trends of central tail quantiles with those of intermediate or extreme tail quantiles. The FTVM-EoT model provides a meaningful structure to address this challenge by enabling the separate analysis of central and extreme tail quantiles. This approach is analogous to the distinct modeling of mean and volatility in statistical literature. For instance, in time series analysis, ARMA or ARIMA models are used to model the conditional mean, while ARCH or GARCH models are employed to model conditional variance. Similarly, in the FTVM-EoT model, \(\mathcal{H}_{0,\tau^*}(\mathcal{I}_{i,t})\) captures the location shift for intermediate and extreme tail quantiles of \(\tilde{U}_{i,t}\left(\cdot \mid \mathcal{I}_{i,t}, f_{0t}, l_{0i}\right)\). On the other hand, \(l_{0i}^\top f_{0t} {U}_{i,t}(\cdot)\) models the excess over \(\mathcal{H}_{0,\tau^*}(\mathcal{I}_{i,t})\) of tail quantiles, where \(l_{0i}^\top f_{0t}\) serves as the scale parameter and \({U}_{i,t}(\cdot)\) determines the tail behavior of the distribution. This clear separation enhances interpretability and facilitates the diagnosis of model components, such as identifying the central tail quantile structure.

Moreover, the FTVM-EoT approach provides a robust framework for estimating tail quantiles. 
Numerous studies have established their statistical properties for central tail quantile estimators. 
For instance, \cite{ando2020quantile} established an error bound of \(O_p(\sqrt{\log{N}/T} \vee \sqrt{\log{T}/N})\) for central tail quantile estimator derived from QFM and QRIFE.
The FTVM-EoT model leverages these mature results to achieve a more accurate estimation of intermediate and extreme tail quantile. 
We demonstrate in Assumption \ref{ass:liploc} and Proposition \ref{cor:qadj} that 
by using these central tail quantile estimators as a foundation, the FTVM-EoT model enhances the robustness and reliability of extreme tail quantile estimation.
Thus, the problem discussed in Example \ref{ex:LSSM} is handled by FTVM-EoT method.

To proceed, let the estimator of \(\mathcal{H}_{0,\tau^*}\) be denoted as \(\hat{\mathcal{H}}_{\tau^*}\). We define \(\hat{U}_{adj}(NT/k)\) as the \(k\)-th largest order statistic of the adjusted sequence \(\{Y_{i,t} - \hat{\mathcal{H}}_{\tau^*}(\mathcal{I}_{i,t})\}_{1 \le i \le N, 1 \le t \le T}\). The adjusted Hill estimator is then given by:
\[
\hat{\gamma}_{adj} = 
k^{-1} \sum_{i=0}^{k} \log\{\hat{U}_{adj}(NT/i)\}  
- \log\{\hat{U}_{adj}(NT/k)\}.
\]
Additionally, the adjusted KS statistic is defined as:
\begin{align*}
 \operatorname{KS}_{adj} := & \sup_{0 \le s \le 1} \sqrt{k} \,\, \Bigg| 
 \left\{ \frac{1}{k} \sum_{t=1}^{\lfloor T s \rfloor} \sum_{i=1}^{N} 
 \mathbf{1}\left( Y_{i,t}  - \hat{\mathcal{H}}_{\tau^*}(\mathcal{I}_{i,t})  \ge \hat{U}\left({NT}/{k}\right) \right) \right\} \\
   & +
 \left\{ \frac{1}{k} \sum_{i=1}^{\lfloor NTs \rfloor - N\lfloor T s \rfloor}
 \mathbf{1}\left( Y_{i,\lfloor Ts \rfloor + 1}  - \hat{\mathcal{H}}_{\tau^*}(\mathcal{I}_{i,\lfloor Ts \rfloor + 1}) \ge 
 \hat{U}\left({NT}/{k}\right) \right) 
 \right\} - s
 \Bigg|.
\end{align*}

We propose Algorithm \ref{alg:TQFM3} as a template for applying the FTVM in conjunction with other statistical methods.

\begin{algorithm}[ht]
   \renewcommand{\algorithmicrequire}{\textbf{Input:}}
   \renewcommand{\algorithmicensure}{\textbf{Output:}}
   \caption{Estimation of FTVM-EoT Method}
   \begin{algorithmic}[1]\label{alg:TQFM3}
      \REQUIRE The data \(Y_{i,t}\)
      \STATE Estimate \(\hat{\mathcal{H}}_{\tau^*}\) satisfying Assumption \ref{ass:liploc}.
      \STATE Estimate the adjusted intermediate tail quantile \(\hat{U}_{adj}(NT/k)\). 
      \STATE Conduct the hypothesis test \(H_0\) using \(\operatorname{KS}_{adj}\). 
      \IF{\(H_0\) is not rejected}
      \STATE \textbf{Output:} The tail quantile factorization 
        \(\hat{\mathcal{H}}_{\tau^*}(\mathcal{I}_{i,t}) + \hat{U}_{adj}(NT/k)\).
      \ELSE
         \STATE Use \(\hat{r}^{adj}_{\mathrm{IC}}\) to determine the optimal number of factors for \(\{Y_{i,t} - \hat{\mathcal{H}}_{\tau^*}(\mathcal{I}_{i,t})\}_{1 \le i \le N, 1 \le t \le T}\).
         \STATE Estimate \(\hat{L}_{N,\hat{r}^{adj}_{\mathrm{IC}}}, \hat{F}_{T,\hat{r}^{adj}_{\mathrm{IC}}}\) using Algorithm \ref{alg:TQFM2} on the data \(\{Y_{i,t} - \hat{\mathcal{H}}_{\tau^*}(\mathcal{I}_{i,t})\}_{1 \le i \le N, 1 \le t \le T}\).
         \STATE \textbf{Output:} The tail quantile factorization 
         \(\hat{\mathcal{H}}_{\tau^*}(\mathcal{I}_{i,t}) + \hat{L}_{N,\hat{r}^{adj}_{\mathrm{IC}}}^\top \hat{F}_{T,\hat{r}^{adj}_{\mathrm{IC}}} \hat{U}_{adj}(NT/k)\).
      \ENDIF       
   \end{algorithmic}
\end{algorithm}

In our research, we do not analyze the convergence of \(\hat{\mathcal{H}}_{\tau^*}\), but instead assume its convergence as a condition.
\begin{ass}\label{ass:liploc}
 The estimator \(\hat{\mathcal{H}}_{\tau^*}\) satisfies that as \( N \), \( T \to \infty \),
   \begin{equation}
 \max_{1 \le i \le N, 1 \le t \le T} \frac{|\hat{\mathcal{H}}_{\tau^*}(\mathcal{I}_{i,t}) - \mathcal{H}_{0,\tau^*}(\mathcal{I}_{i,t}) |}{U(NT/k)} = O_p(B_{N,T} k^{-1/2}),
   \end{equation}where $B_{N,T} \to 0$ as \( N \), \( T \to \infty \).
\end{ass}

We next present the following proposition. The proposition reveals that under Assumption \ref{ass:liploc}, the error caused by $\hat{\mathcal{H}}_{\tau^*}(\mathcal{I}_{i,t}) - {\mathcal{H}_{0,\tau^*}}(\mathcal{I}_{i,t})$ has no essential influence on the estimation of intermediate tail quantile and the procedure of validation test.

\begin{prop}\label{cor:qadj}
 Suppose the estimated threshold model $\hat{\mathcal{H}}_{\tau^*}(\mathcal{I}_{i,t})$ at a central tail quantile level $\tau^*$ satisfies Assumption \ref{ass:liploc} and the excess sequence \(\{Y_{i,t} - \mathcal{H}_{0,\tau^*}(\mathcal{I}_{i,t})\}\) satisfies the FTVM in \eqref{eq:ftvm} with Assumptions \ref{ass: identify}-\ref{ass:intermediate}. as \( N \), \( T \to \infty \),
   \begin{enumerate}
      \item for the $k$-th biggest order statistic $\hat{U}_{adj}(NT/k)$, it holds that
       \begin{equation*}
 \sqrt{k} \left(\frac{\hat{U}_{adj}(NT/k)}{U(NT/k)} - 1\right) \leadsto N(0, \gamma^2).
      \end{equation*}
      \item for the Hill estimator $\hat{\gamma}_{adj}$, it holds that 
      $
 \sqrt{k} \left(\hat{\gamma}_{adj} - \gamma\right) \leadsto N(0, \gamma^2).
      $
      \item for the KS statistic $\operatorname{KS}_{adj}$, under \(H_0\), there exists a Brownian Bridge \(B\) such that as \( N \), \( T \to \infty \),
      $
 \operatorname{KS}_{adj} \leadsto \sup_{0 \le s \le 1} |B(s)|.
      $
   \end{enumerate}
\end{prop}
\begin{proof}
 We verify that \(\{Y_{i,t} - \hat{\mathcal{H}}_{\tau^*}(\mathcal{I}_{i,t})\}\) satisfies the conditions in Assumptions \ref{ass: identify}-\ref{ass:intermediate}. Denote 
   \[
 \frac{Y_{i,t} - \hat{\mathcal{H}}_{\tau^*}(\mathcal{I}_{i,t})}{l_{0i}^\top f_{0t}} = U_{i,t}(V_{i,t}^{-1}) + R_{i,t},
   \]
 where \(R_{i,t}\) satisfies \(\max_{1 \le i \le N, 1 \le t \le T} |R_{i,t}| = O_p( B_{N,T} U(NT/k)k^{-1/2})\) as \(N, T \to \infty\). 
 We obtain that with probability tending to 1,
   \[
 \left|\frac{U_{i,t}(x) + R_{i,t}}{U(x)} - 1 \right| \le \left|\frac{U_{i,t}(x)}{U(x)} - 1 \right| + \frac{|R_{i,t}|}{U(x)} = O(A_{N,T}(x)) + O\left( B_{N,T} k^{-1/2} \right).
   \]
 The last step follows by $U({NT/k}) / U(x)$ is totally bounded for $x > NT/k(2M^{1/\gamma})$ as \( N \), \( T \to \infty \).
 Thus, the conditions of Assumptions \ref{ass: identify}-\ref{ass:intermediate} are satisfied.
\end{proof}

To summarize, we state the following convergence result of the estimated intermediate tail quantiles, extreme tail quantiles, and the consistency of the factor numbers.
Denote \(\hat{r}^{adj}_{\mathrm{IC}}\) as the optimal number of factors by applying \eqref{eq:ric} on the data \(\{Y_{i,t} - \hat{\mathcal{H}}_{\tau^*}(\mathcal{I}_{i,t})\}_{1 \le i \le N, 1 \le t \le T}\).
Denote the MSRE metric for the tail quantile level $\tau$ as 
\begin{align}
   &\, \operatorname{MSRE}^{EoTM}_{\tau} ({\mathcal{H}}, L_{N,r}, F_{T,r}, \Lambda) \nonumber \\
 := &
 {\frac{1}{NT}
   \sum_{i=1}^{N} \sum_{t=1}^{T} \left(
 {\frac{\mathcal{H}(\mathcal{I}_{i,t}) +  {l}_{i,r}^\top {f}_{t,r} \Lambda}{U(\tau^{-1})}
 } - \frac{ \tilde{U}_{i,t}\left({\tau}^{-1} \mid \mathcal{I}_{i,t}, f_{0t}, l_{0i} \right) }{{U(\tau^{-1})}}
 \right)^2}. \label{eq:MSREEOT}
   \end{align}

\begin{cor}\label{thm:qfa}
 Under the conditions of Proposition \ref{cor:qadj}, 
 suppose $p_{N,T}$ is an extreme tail quantile level such that \( NTp_{N,T} = o(k) \) and \( \log(NTp_{N,T}) = o(\sqrt{k}) \) and $P_{N,T}$ is a threshold such that
    \(P_{N,T} (k /(N+T)) \to \infty\) and \(P_{N,T} \to 0\) 
 as \( N \to \infty \) and \( T \to \infty \). Then, as \( N \to \infty \) and \( T \to \infty \),
   \begin{enumerate}
      \item for the intermediate tail quantile factorization, it holds that
   \begin{align*}
 \operatorname{MSRE}^{EoTM}_{k/NT} \left(\hat{\mathcal{H}}_{\tau^*}, \hat{L}_{N,r}, \hat{F}_{T,r}, \hat{U}_{adj}\left(\frac{NT}{k}\right) \right) = O_p\left( \frac{N+T}{k} \right).
    \end{align*}
    \item for the extreme tail quantile factorization, it holds that
    \begin{align*}
 & \operatorname{MSRE}^{EoTM}_{p_{N,T}} \left(\hat{\mathcal{H}}_{\tau^*}, \hat{L}_{N,r}, \hat{F}_{T,r}, \hat{U}_{adj}\left(\frac{NT}{k}\right) 
 \left(\frac{k}{NTp_{N,T}}\right)^{\hat{\gamma}_{adj}} \right) 
 \\
 = & \, O_p\left(\frac{N+T}{k} \vee \frac{\log^2(k/(NT p_{N,T}))}{{k}}\right).
    \end{align*}
    \item for the estimator of factor numbers, it holds that
   $
 P(\hat{r}^{adj}_{\mathrm{IC}} = r) \to 1.
   $
   \end{enumerate}   
\end{cor}

\section{Simulation}

In this section, we conduct simulation studies to evaluate the performance of the proposed methods under various data generation processes (DGPs). The simulations are designed to assess the accuracy and robustness of the FTVM, the FTVM-EoT method, and other benchmark methods, such as QFM and QRIFE.
We focus on both intermediate and extreme tail quantile estimation problems, particularly in scenarios involving heavy-tailed distributions. Key performance metrics, such as MSREs defined in \eqref{eq:MSREdef} and \eqref{eq:MSREEOT}, are used to compare the methods across different sample sizes, quantile levels, and tail indices.

\subsection{Data Generation Process}

In this subsection, we introduce the DGPs applied in this paper. 
The DGPs are carefully constructed to reflect serial correlation and multi-factor models. 
We first introduce the DGPs for the FTVM. The simulated data follows the model \(Y_{i,t} = l_{i}^\top f_{t} u_{i,t} b_{i,t}\), where the specifications for \(l_{i}\), \(f_{t}\), \(u_{i,t}\), and \(b_{i,t}\) are detailed below. The term \(u_{i,t}\) is generated independently from a Pareto distribution with a tail quantile function given by \(x^{1/\lambda}\), where \(\lambda\) is the shape parameter. We consider \(\lambda = 1, 2\), and \(3\), corresponding to cases where \(\gamma = 1, 1/2\) and \(1/3\). The term \(b_{i,t}\) is also generated independently from  Rademacher distribution.
{ This generation models risks of returns in financial markets}, where extreme tail behavior is prevalent, and returns can be either positive or negative.
For each DGP that generates loadings and factors, we define the reference tail quantile function as
$
U(x) = c \cdot (x/2)^{1/\lambda}$, for $x > 2$,
where \(c := c(\lambda, \operatorname{DGP}) = \lim_{N, T \to \infty}\{(NT)^{-1} \sum_{i=1}^{N} \sum_{t=1}^{T} (l_{i}^\top f_{t})^{\lambda}\}^{1/\lambda}\). The constant \(c\) is estimated as the finite sample mean of \(\{(NT)^{-1} \sum_{i=1}^{N} \sum_{t=1}^{T} (l_{i}^\top f_{t})^{\lambda}\}^{1/\lambda}\). We consider cases where \(N = 50, 100, 200\) and \(T = 50, 100, 200\).
We generate \(\operatorname{Beta}(a,b)\) following the density function of the beta distribution with shape parameters \(a\) and \(b\).
Next, we describe the generation of \(l_i\) and \(f_t\) for the three DGPs:
\begin{description}[nolistsep]
    \item[DGP1] A single-factor model where \(l_i\) is generated as a shifted \(\operatorname{Beta}(1,1)\) random variable, and \(f_t\) follows an AR(1) process with \(\operatorname{Beta}(1,1)\) innovations and a constant shift:  
\begin{flalign*}
 \left\{  
  \begin{aligned}
    & l_i = 0.5 + \epsilon_i,  &  \epsilon_i \sim \operatorname{Beta}(1,1), & \text{ for } 1 \leq i \leq N, \\
    & f_t = 0.4 f_{t-1} + \varepsilon_t + 0.3, & \varepsilon_t \sim \operatorname{Beta}(1,1), & \text{ for } 1 \leq t \leq T.
  \end{aligned}
 \right.
\end{flalign*}
 Here, \(\epsilon_i\) and \(\varepsilon_t\) are i.i.d. random variables. DGP1 is designed to evaluate the performance of the proposed method when factors exhibit serial correlation. 
  \item[DGP2] A two-factor model where \(l_i\) is a two-dimensional vector with each component sampled as a shifted \(\operatorname{Beta}(0.5,0.5)\) random variable, and \(f_t\) is a shifted vector autoregressive process with \(\operatorname{Beta}(0.5,0.5)\) innovations:
  \begin{flalign*}
 \left\{  
  \begin{aligned}
    & l_i = 0.5 + \begin{bmatrix}
       \epsilon_{1,i} \\ \epsilon_{2,i}
    \end{bmatrix},  &  \epsilon_{j,i} \sim \operatorname{Beta}(0.5,0.5), & \text{ for } 1 \leq i \leq N, \, j = 1,2, \\
    & f_t = \begin{bmatrix}
    0.4 & 0 \\ 0 & 0.2
    \end{bmatrix} f_{t-1} +  \begin{bmatrix}
       \varepsilon_{1,t} \\ \varepsilon_{2,t}
    \end{bmatrix} + \begin{bmatrix}
       0.3 \\ 0.4
    \end{bmatrix}, &  \varepsilon_{j,t} \sim \operatorname{Beta}(0.5,0.5), & \text{ for } 1 \leq t \leq T, \, j = 1,2.
  \end{aligned}
 \right.
\end{flalign*}
 Here, \(\varepsilon_{j,t}\) and \(\epsilon_{j,i}\) are i.i.d. random variables.
  \item[DGP3]  This data generation process involves solving the optimization problem to maximize \(\sigma_{N2}\) by the following steps:
  \begin{enumerate}[nolistsep]
    \item Generate \(\epsilon_{j,i} \sim \operatorname{Beta}(0.5, 0.5)\) for \(1 \leq i \leq N\), \(j = 1, 2\), and \(\varepsilon_{j,t} \sim \operatorname{Beta}(0.5, 0.5)\) for \(1 \leq t \leq T\), \(j = 1, 2\), where \(\varepsilon_{j,t}\) and \(\epsilon_{j,i}\) are i.i.d. random variables.
    \item Perform singular value decomposition (SVD) on the matrix:
    \[
    0.5 + 
    \begin{bmatrix}
       \epsilon_{1,1}, \ldots , \epsilon_{1,N} \\ \epsilon_{2,1}, \ldots , \epsilon_{2,N}
    \end{bmatrix}^\top 
    \begin{bmatrix}
       \varepsilon_{1,1}, \ldots, \varepsilon_{1,T} \\ \varepsilon_{2,1}, \ldots, \varepsilon_{2,T}
    \end{bmatrix}
 = \mathcal{V}^\top \mathcal{D} \mathcal{W},
    \]
 where \(\mathcal{V} = [{v}_{1}, \ldots, {v}_{N}] \in \mathbb{R}^{2 \times N}\) and \(\mathcal{W} = ({w}_{1}, \ldots, {w}_{T}) \in \mathbb{R}^{2 \times T}\).
    \item Solve for \((\varsigma_1, \varsigma_2)\) by optimizing:
    \[
 (\varsigma_1, \varsigma_2) = \arg \max \varsigma_2, \quad \text{s.t. } \varsigma_1 \geq \varsigma_2, \, 0.1 \leq v_i \operatorname{diag}(\varsigma_1, \varsigma_2) w_t^\top \leq 5.
    \]
    \item Return \([l_1, \ldots, l_N] = \mathcal{V}^\top \operatorname{diag}(\varsigma_1, \varsigma_2) / \sqrt{T}\) and \([f_1, \ldots, f_T] = \sqrt{T} \mathcal{W}^\top\).
  \end{enumerate}

\end{description}

\begin{rem}
We observe that for DGP2, \(\sigma_{N2}\) is significantly smaller than \((N+T)/k\), even when \(N=200\) and \(T=200\). Table \ref{tab:faclod} reports the minimum, median, and maximum values of \(\sigma_{N1}\) and \(\sigma_{N2}\) for DGP2 and DGP3. The small value of \(\sigma_{N2}\) makes it challenging to distinguish the corresponding factors and loadings. As reflected in the simulation results, factor selection methods tend to favor single-factor models and degenerate FTVM.

To address this issue, we introduce DGP3 to evaluate the validity of factor selection methods. In DGP3, \(\sigma_{N2}\) is comparable to \((N+T)/k\), particularly when \(k = 0.1 NT\) and \((N, T) = (200, 200)\). However, for smaller sample sizes, such as \((N, T) = (50, 50)\), \(\sigma_{N2}\) in DGP3 is approximately 10 times smaller than \((N+T)/k\). This highlights the increased difficulty of selecting the appropriate number of factors in small sample settings.

\begin{table}[htbp]
    \centering
    \caption{Simulation results for DGPs 2 and 3, presenting the minimum, median, and maximum values of $\sigma_{N1}$ and $\sigma_{N2}$, and $(N + T)/k$ across varying sample sizes $(N, T)$.
   }
    \resizebox{\linewidth}{!}{
       \begin{tabular}{rrrcrrrrrrrrrrrr}
           \toprule
            &   &   &   &   & \multicolumn{3}{c}{$\sigma_{N1}$} &   & \multicolumn{3}{c}{$\sigma_{N2}$} &   & \multicolumn{2}{c}{$(N+T)/k$} &  \\
         \cmidrule{6-8}\cmidrule{10-12}\cmidrule{14-15}    & \multicolumn{1}{c}{DGP} & \multicolumn{1}{c}{$\lambda$} & (N,T) &   & \multicolumn{1}{c}{Min.} & \multicolumn{1}{c}{Median} & \multicolumn{1}{c}{Max.} &   & \multicolumn{1}{c}{Min.} & \multicolumn{1}{c}{Median} & \multicolumn{1}{c}{Max.} &   & \multicolumn{1}{c}{$k = 0.1 NT $} & \multicolumn{1}{c}{$k= 0.05 NT$} &  \\
         \cmidrule{2-15}    & DGP2 & 1  & (50, 50) &   & 1.062 & 1.110 & 1.165 &   & 0.001 & 0.003 & 0.006 &   & 0.400 & 0.800 &  \\
            &   &   & (100, 100) &   & 1.077 & 1.111 & 1.144 &   & 0.001 & 0.003 & 0.005 &   & 0.200 & 0.400 &  \\
            &   &   & (200, 200) &   & 1.084 & 1.111 & 1.143 &   & 0.002 & 0.003 & 0.004 &   & 0.100 & 0.200 &  \\
         \cmidrule{4-15}    & DGP3 & 3  & (50, 50) &   & 0.765 & 0.820 & 0.865 &   & 0.018 & 0.044 & 0.090 &   & 0.400 & 0.800 &  \\
            &   &   & (100, 100) &   & 0.787 & 0.823 & 0.853 &   & 0.023 & 0.042 & 0.074 &   & 0.200 & 0.400 &  \\
            &   &   & (200, 200) &   & 0.799 & 0.824 & 0.843 &   & 0.030 & 0.041 & 0.061 &   & 0.100 & 0.200 &  \\
           \bottomrule
           \end{tabular}%
 }
    \label{tab:faclod}%
   \end{table}%

\end{rem}



We then describe the data generation processes for the FTVM-EoT models. Especially, we choose QFM and QRIFE to serve as the threshold models at a central quantile level as well as the benchmark models without the enhancement of FTVM.
\begin{description}[nolistsep]
    \item[DGP4] 
 DGP4 is defined as \( Y_{i,t} = a_i b_t + l_i f_t u_{i,t} \), where \( a_i \), \( b_t \), \( l_i \), and \( f_t \) are real-valued, and \( u_{i,t} \) follows a Student-t distribution with degrees of freedom \(\lambda\). The loadings and factors \( l_i \) and \( f_t \) are generated from DGP1, while \( a_i \) and \( b_t \) are generated as follows:  
    \begin{flalign*}
 \left\{ 
    \begin{aligned}
 a_i &\sim N(1, 1), & \text{for } 1 \leq i \leq N, \\
 b_t &= 0.6 b_{t-1} + \eta_t,  & \eta_t \sim N(1, 1), \quad \text{for } 1 \leq t \leq T.
    \end{aligned} 
 \right.
    \end{flalign*}
 This model is a special case of the QFM proposed by \cite{Chen2021}.
    \item[DGP5]
 DGP5 is defined as \( Y_{i,t} = \textbf{x}_{i,t}^\top b_i + l_i f_t u_{i,t} \), where \( l_i \) and \( f_t \) are generated from DGP1, and \( u_{i,t} \) follows a Student-t distribution with degrees of freedom \(\lambda\). The covariate \( \textbf{x}_{i,t}, b_{i} \in \mathbb{R}^2 \) are two-dimensional vectors defined as:
    \[
 \left\{ 
    \begin{aligned}
 x_{i,t} & = \begin{bmatrix}
         \eta_{i,t,1} + 0.2 f_t^2 + 0.8 l_i^2 \\
         \eta_{i,t,2}
       \end{bmatrix}, &  \eta_{i,t,1}, \eta_{i,t,2} \sim N(1, 1), \text{ for } 1 \leq i \leq N, 1 \leq t \leq T, \\
 b_{i,t} & =  - \frac{1}{2} +  \begin{bmatrix}
         \zeta_{i,1}  \\
         \zeta_{i,2}
       \end{bmatrix}, &  \zeta_{i,1}, \zeta_{i,2} \sim \operatorname{Beta}(1, 1), \text{ for } 1 \leq i \leq N.
    \end{aligned} 
 \right.
    \]
 This model is a special case of the QRIFE proposed by \cite{ando2020quantile}.
    
\end{description}

\subsection{Simulation Results for Factor Tail Volatility Model}\label{subsec:simFTVM}
In this subsection, we present the simulation results for the FTVM. The analysis is divided into three parts. First, we investigate the MSREs of the FTVM at intermediate tail quantile levels, comparing its performance with degenerate FTVMs and alternative factor numbers. Second, we explore the impact of the tuning parameter \(M\) on the model's accuracy and discuss the implications of overfitting when \(M\) is excessively large. Additionally, we evaluate the effectiveness of model validation and factor selection methods in identifying the appropriate number of factors.



\subsubsection{Mean Squared Relative Errors of Intermediate Tail Quantiles}
\begin{table}[htbp]
   
    \centering
    \caption{Simulated MSREs at $k/NT = 0.1, 0.05$ across varying sample sizes $(N, T)$ for different DGPs  and $\lambda$.
 We calculate $\operatorname{MSRE}_{k/NT}(\mathbf{1}^{N}, \mathbf{1}^{T}, \hat{U}(NT/k) )$ for the degenerate FTVM, corresponding to `$r=0$' in the table.
  $\operatorname{MSRE}_{k/NT}(\hat{L}_{N,r}, \hat{F}_{T,r}, \hat{U}(NT/k) )$ is calculated for the FTVM with $r = 1, 2, 3$. In all the experiments, we set $M = 1.6$ and $m = 0.1$. For each experiment, we replicate 1000 times and report the average MSREs.}
    
    \resizebox{\linewidth}{!}{
       \begin{tabular}{rrrccrrrrrrrrrr}
           \toprule
            &   &   &   &   & \multicolumn{4}{c}{$\operatorname{MSRE}_{0.1}$ $(\times 10^{-3})$} &   & \multicolumn{4}{c}{$\operatorname{MSRE}_{0.05}$ $(\times 10^{-3})$} &  \\
         \cmidrule{6-9}\cmidrule{11-14}    & \multicolumn{1}{l}{DGP} & \multicolumn{1}{l}{$\lambda$} & (N,T) &   & \multicolumn{1}{c}{$r=0$} & \multicolumn{1}{c}{$r=1$} & \multicolumn{1}{c}{$r=2$} & \multicolumn{1}{c}{$r=3$} &   & \multicolumn{1}{c}{$r=0$} & \multicolumn{1}{c}{$r=1$} & \multicolumn{1}{c}{$r=2$} & \multicolumn{1}{c}{$r=3$} &  \\
         \cmidrule{2-14}    & DGP1 & 2  & (50, 50) &   & 129.9 & \textbf{57.3} & 92.2 & 111.0 &   & 130.2 & \textbf{92.3} & 132.0 & 151.1 &  \\
            &   &   & (50, 100) &   & 129.0 & \textbf{49.2} & 78.5 & 96.6 &   & 129.1 & \textbf{83.3} & 119.2 & 136.6 &  \\
            &   &   & (100, 100) &   & 129.3 & \textbf{33.9} & 60.5 & 76.1 &   & 129.4 & \textbf{59.8} & 96.4 & 113.4 &  \\
            &   &   & (200, 200) &   & 129.7 & \textbf{19.8} & 37.1 & 48.2 &   & 129.7 & \textbf{35.2} & 63.4 & 77.7 &  \\
         \cmidrule{4-14}    & DGP2 & 1  & (50, 50) &   & \textbf{117.0} & 154.1 & 213.1 & 812.5 &   & \textbf{120.5} & 217.1 & 280.6 & 812.1 &  \\
            &   &   & (50, 100) &   & \textbf{115.2} & 136.5 & 194.4 & 809.7 &   & \textbf{116.9} & 196.5 & 264.0 & 810.0 &  \\
            &   &   & (100, 100) &   & 115.0 & \textbf{103.5} & 160.4 & 809.3 &   & \textbf{115.8} & {157.4} & 225.5 & 809.4 &  \\
            &   &   & (200, 200) &   & 114.7 & \textbf{66.4} & 112.6 & 809.3 &   & 114.9 & \textbf{107.4} & 168.0 & 809.6 &  \\
         \cmidrule{4-14}    & DGP3 & 3  & (50, 50) &   & 171.0 & 81.4 & \textbf{71.9} & 82.4 &   & 171.4 & 97.3 & \textbf{115.0} & 124.0 &  \\
            &   &   & (50, 100) &   & 169.8 & 76.6 & \textbf{49.4} & 59.8 &   & 170.4 & 92.6 & \textbf{102.5} & 111.9 &  \\
            &   &   & (100, 100) &   & 169.2 & 66.6 & \textbf{27.4} & 36.6 &   & 169.9 & 79.1 & \textbf{74.8} & 81.3 &  \\
            &   &   & (200, 200) &   & 168.2 & 58.0 & \textbf{14.5} & 19.8 &   & 169.0 & 65.6 & \textbf{30.4} & 37.9 &  \\
           \bottomrule
           \end{tabular}%
 }
    \label{tab:CV}%
   \end{table}%
Simulated MSREs are presented in Table \ref{tab:CV}. Across all values of \(\lambda\), the MSREs of the factor models decrease as the sample size \((N, T)\) increases. In contrast, the MSREs of the degenerate FTVM remain relatively constant. The MSREs are smaller when \(k = 0.1 NT\).

For DGP1, the FTVM with \(r = 1\) consistently outperforms the degenerate FTVM and other factor models. Interestingly, when \((N, T) = (50, 50)\) and \((50, 100)\) with \(k = 0.05 NT\), the degenerate FTVM achieves smaller MSREs compared to factor models with \(r = 2,3\).

For DGP2, the degenerate FTVM outperforms the FTVM when \((N, T) = (50, 50)\) and \((50, 100)\). The FTVM with higher factor numbers (\(r = 2, 3\)) performs poorly, with significantly larger MSREs (e.g., \(r = 3\) reaches \(812.1 \times 10^{-3}\) at \((N, T) = (50, 50)\) and \(k = 0.05 NT\)). These results suggest that the FTVM is not well-suited for DGP2 in small sample size settings, particularly when \(N < 100\) and \(T < 100\). However, for larger sample sizes \((N, T) = (100, 100)\) and \((200, 200)\), the FTVM with \(r = 1\) demonstrates better performance. This is consistent with the small \(\sigma_{N2}\) values reported in Table \ref{tab:faclod}, where the uncertainty introduced by solving the factor model outweighs the benefits of estimating a two-factor model if \(\sigma_{N2}\) is too small relative to \((N+T)/k\). Given the slow convergence rate of \((N + T) / k\) to 0 as \(N, T \to \infty\), we recommend \(r = 1\) as a practical choice for DGP2, especially when data with large sample size is unavailable.

For DGP3, the FTVM with \(r = 2\) achieves the lowest MSREs at \((N, T) = (200, 200)\). For data with smaller sample size (e.g., \((N, T) = (50, 50)\)), the optimal number of factors depends on the ratio \(k / NT\), with \(r = 1\) or \(r = 2\) performing best in different scenarios. This pattern aligns with the results in Table \ref{tab:faclod}, where \(\sigma_{N2}\) is comparable to \((N + T) / k\). 

\subsubsection{Mean Squared Relative Errors Under Different Upper Bound M}
   \begin{table}[t]
    \centering
    \caption{Simulated MSREs at $k/NT = 0.1, 0.05$ with different $M$. In the experiment, we generate $l_i$ anf $f_t$ from DGP3. We analyze MSREs of FTVM under different $\lambda$ across different $M$. In the experiment, $m$ is set as $0.1$. For each experiment, we replicate for 100 times and return the average MSREs.}
    \resizebox{\linewidth}{!}{ 
       \begin{tabular}{rcrrrrrrrrrrrrrrrr}
           \toprule
            &   &   &   & \multicolumn{6}{c}{$\operatorname{MSRE}_{0.1}$ $(\times 10^{-3})$} &   & \multicolumn{6}{c}{$\operatorname{MSRE}_{0.05}$ $(\times 10^{-3})$} &  \\
            \cmidrule{5-10}\cmidrule{12-17}    & (N,T) & \multicolumn{1}{c}{$\lambda$} &   & \multicolumn{1}{c}{1} & \multicolumn{1}{c}{1.3} & \multicolumn{1}{c}{1.6} & \multicolumn{1}{c}{2} & \multicolumn{1}{c}{6} & \multicolumn{1}{c}{32} &   & \multicolumn{1}{c}{1} & \multicolumn{1}{c}{1.3} & \multicolumn{1}{c}{1.6} & \multicolumn{1}{c}{2} & \multicolumn{1}{c}{6} & \multicolumn{1}{c}{32} &  \\
            \cmidrule{2-17}    & (50, 50) & 1  &   & 255 & \textbf{243} & 272 & 345 & 1514 & 8919 &   & 324 & \textbf{313} & 346 & 428 & 2193 & 13042 &  \\
                &   & 2  &   & 131 & \textbf{115} & 129 & 162 & 393 & 552 &   & 158 & \textbf{153} & 174 & 220 & 665 & 1336 &  \\
                &   & 3  &   & 86 & \textbf{63} & 72 & 95 & 154 & 155 &   & 102 & \textbf{98} & 114 & 143 & 280 & 321 &  \\
            \cmidrule{3-17}    & (100, 100) & 1  &   & 215 & \textbf{195} & 213 & 265 & 915 & 3518 &   & 266 & \textbf{260} & 286 & 359 & 1508 & 9314 &  \\
                &   & 2  &   & 111 & \textbf{73} & 72 & 93 & 194 & 208 &   & 133 & \textbf{119} & 134 & 165 & 400 & 624 &  \\
                &   & 3  &   & 69 & 33 & \textbf{28} & 36 & 54 & 53 &   & 86 & \textbf{66} & 74 & 96 & 161 & 165 &  \\
            \cmidrule{3-17}    & (200, 200) & 1  &   & 424 & \textbf{151} & 157 & 190 & 490 & 851 &   & 221 & 270 & \textbf{224} & 277 & 936 & 3432 &  \\
                &   & 2  &   & 95 & 47 & \textbf{34} & 39 & 69 & 67 &   & 114 & \textbf{81} & 82 & 108 & 213 & 228 &  \\
                &   & 3  &   & 62 & 23 & \textbf{15} & 16 & 19 & 20 &   & 70 & 35 & \textbf{31} & 38 & 62 & 61 &  \\
              \bottomrule
              \end{tabular}%
 }
    \label{tab:M}%
   \end{table}%

To evaluate the performance of the FTVM under varying values of \(M\), we report the MSREs in Table~\ref{tab:M}. 
The MSREs reach their minimum at \(M = 1.3\) or \(M = 1.6\), depending on \((N, T, k)\) and \(\lambda\). Specifically, when \((N, T) = (200, 200)\), \(\lambda = 1, 2\) favors \(M = 1.3\), while \(\lambda = 3\) favors \(M = 1.6\). These results suggest that the optimal choice of \(M\) depends on both the dimensionality of the data and the extreme value index \(\gamma = 1/\lambda\).
    


\begin{figure}[htbp]
    \centering
      
       
\includegraphics[width=.7\linewidth]{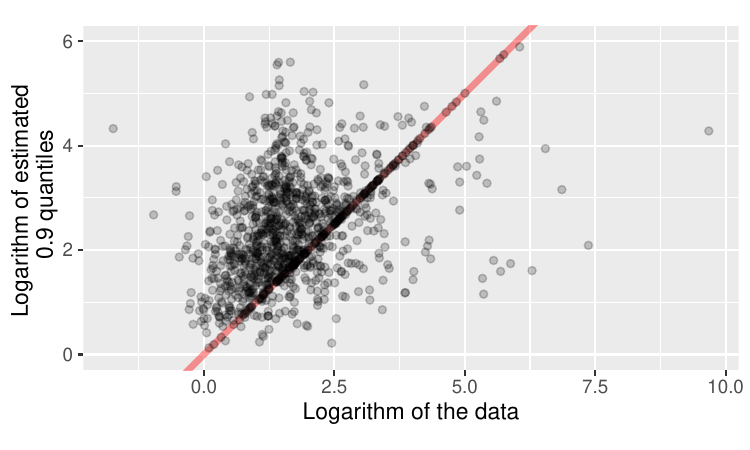}
    \caption{Scatter plot of $(\log(Y_{i,t}), \log(\hat{l}_{i,2}^\top \hat{f}_{t,2} \hat{U}(10) ))$, where $Y_{i,t}$ is generated from DGP3 with $\lambda = 1$, $(N,T) = (50,50)$, and $\hat{l}_{i,2}^\top$, $\hat{f}_{t,2}^\top$ are estimated by FTVM with $r=2$, $m = 0.1$ and $M =32$
 The  solid line represents the line $y = x$.} 
    \label{fig:scatter}
\end{figure}

 Notably, the performance of the FTVM worsens when \(M\) becomes excessively large. For instance, when \(M = 32\) with \((N, T) = (50, 50)\), \(\operatorname{MSRE}_{0.1}\) increases significantly to \(8.919\). 
 A plausible explanation for this phenomenon is that an overly large \(M\) causes the FTVM to overfit the observed data rather than accurately estimating the intermediate tail quantiles. 
  Evidence from Figure~\ref{fig:scatter} shows that when \(M = 32\), some scatter points align precisely with the identity line \(y = x\), suggesting that FTVM overfits individual observations rather than capturing the underlying quantile relationship.
    
    

\subsubsection{Model Validation and Factor Selection}

\begin{table}[t]
    \centering
    \caption{ Simulated \textit{Rejection frequency} (RF) of the testing $H_0$, 
 estimated $\hat{r}_{\text{IC}}$, and the frequency that $\hat{r}_{\text{IC}} = r$ (PE). For each case, the experiments are replicated 1000 times. $M = 1.6$ and $m=0.01$ is set in the experiment. 
 }
    \resizebox{.85\linewidth}{!}{
       \begin{tabular}{rrrccrrrrrrrr}
           \toprule
            &   &   &   &   & \multicolumn{3}{c}{$k=0.1NT$} &   & \multicolumn{3}{c}{$k=0.05NT$} &  \\
         \cmidrule{6-8}\cmidrule{10-12}    & \multicolumn{1}{l}{DGP} & \multicolumn{1}{l}{$\lambda$} & (N,T) &   & \multicolumn{1}{c}{RF} & \multicolumn{1}{c}{$\hat{r}_{\text{IC}}$} & \multicolumn{1}{c}{PE} &   & \multicolumn{1}{c}{RF} & \multicolumn{1}{c}{$\hat{r}_{\text{IC}}$} & \multicolumn{1}{c}{PE} &  \\
         \cmidrule{2-12}    & \multicolumn{1}{l}{DGP1} & 2  & (50, 50) &   & 46.1\% & 1.63  & 42.5\% &   & 26.3\% & 3.00  & 0.0\% &  \\
            &   &   & (50, 100) &   & 50.7\% & 1.02  & 98.3\% &   & 29.5\% & 2.83  & 0.0\% &  \\
            &   &   & (100, 100) &   & 78.6\% & 1.00  & 100.0\% &   & 49.3\% & 1.64  & 38.8\% &  \\
            &   &   & (200, 200) &   & 97.1\% & 1.00  & 100.0\% &   & 84.8\% & 1.00  & 100.0\% &  \\
         \cmidrule{4-12}    & \multicolumn{1}{l}{DGP2} & 1  & (50, 50) &   & 11.2\% & 1.00  & 0.0\% &   & 7.4\% & 1.75  & 75.3\% &  \\
            &   &   & (50, 100) &   & 11.4\% & 1.00  & 0.0\% &   & 8.1\% & 1.07  & 6.9\% &  \\
            &   &   & (100, 100) &   & 20.9\% & 1.00  & 0.0\% &   & 11.1\% & 1.00  & 0.0\% &  \\
            &   &   & (200, 200) &   & 44.0\% & 1.00  & 0.0\% &   & 25.2\% & 1.00  & 0.0\% &  \\
         \cmidrule{4-12}    & \multicolumn{1}{l}{DGP3} & 3  & (50, 50) &   & 43.4\% & 2.30  & 66.7\% &   & 24.1\% & 3.00  & 0.5\% &  \\
            &   &   & (50, 100) &   & 47.1\% & 2.03  & 81.8\% &   & 26.6\% & 2.85  & 15.5\% &  \\
            &   &   & (100, 100) &   & 76.9\% & 2.02  & 97.2\% &   & 52.1\% & 2.27  & 66.4\% &  \\
            &   &   & (200, 200) &   & 96.8\% & 2.00  & 99.8\% &   & 80.5\% & 2.04  & 95.0\% &  \\
           \bottomrule
           \end{tabular}%
 }
    \label{tab:selection}%
   \end{table}%
   
We utilize the penalty term \(P_{N,T}\) in \eqref{eq:PPNT} with $c=10$. 
Table \ref{tab:selection} presents the performance of the model validation and factor number estimation methods. The rejection frequency of \(H_0\) increases as \((N, T)\) grow larger. Notably, the rejection frequency is lower when \(k = 0.05 NT\). For DGP1 and DGP3, the rejection frequency approaches 1 when \((N, T) = (200, 200)\), whereas it remains low for DGP2. These findings are consistent with the experimental results in Table \ref{tab:CV}.
Next, we analyze the performance of the estimators of factor numbers. For sufficiently large \((N,T)\), the average estimated factor number \(\hat{r}_ {\operatorname{IC}}\) converges to the true factor numbers for both DGP1 and DGP3, with the correct selection frequency \(\Pm(\hat{r}_{\operatorname{IC}} = r)\) approaching 1. However, in small-dimensional settings, the estimator tends to select over-parameterized FTVM. This phenomenon occurs particularly when the ratio \(k/(N+T)\) is small (e.g., when \((N,T) = (50,50)\), \(k = 0.05NT\), \(k / (N+T) = 2.5\)), thereby violating the conditions required by Theorem~\ref{thm:asymrate}. Considering the results in Table \ref{tab:CV}, which demonstrate that the FTVM performs poorly with large factor numbers, we recommend that model validation is essential in such cases.

\subsection{Simulation Results for QFM-FTVM and QRIFE-FTVM}\label{subsec:simEoTM}

\label{sec:EOTsimulation}

Firstly, we describe the models applied in the simulation experiments.

\begin{description}
  \item[QFM] QFM is implemented directly at intermedaite and extreme tail quantile levels by using the \textit{Iterative Quantile Regression} method introduced in \cite{Chen2021} to estimate the parameters of the quantile factor models. 
  
  \item[QRIFE] QRIFE is implemented at intermedaite and extreme tail quantile levels by using the frequency method introduced in \cite{ando2020quantile}. This method estimates \(\hat{b}_t\), \(\hat{L}_{N,1}\), and \(\hat{F}_{T,1}\) simultaneously.
  
  
  \item[EoTM-0] EoTM-0 estimates the median quantiles of the data using QFM for DGP4 and quantile regression for DGP5, and then applies the degenerate FTVM to estiamte the excess data at intermedaite and extreme tail quantile levels. Intermediate tail quantiles are estimated using \(\hat{U}_{adj}(NT/k)\), while extreme tail quantiles at level \(p_{N,T}\) are estimated using \(\hat{U}_{adj}\left({NT}/{k}\right) \left({k}/({NTp_{N,T}})\right)^{\hat{\gamma}_{adj}}\). This model serves as a benchmark for estimating intermediate and extreme tail quantiles of high-dimensional data.

  \item[EoTM-1] EoTM-1 follows the same approach of EoTM-0 but with a fixed factor number of 1 in FTVM. {The $\mathcal{H}_{0,0.5}$ is estimated using QFM for DGP4 and panel quantile regression for DGP5, respectively.}
    
  \item[FTVM] FTVM is implemented directly to the data with a fixed factor number of 1.
\end{description}

We analyze the performance of the introduced methods in the following subsections. First, we evaluate the performance of the FTVM-EoT models under intermediate tail quantile settings. Second, we assess the methods under extreme tail quantile settings, examining their ability to handle heavy-tailed distributions. 


\subsubsection{Mean Squared Relative Error of Intermediate Tail Quantiles}

We report the MSREs for DGP4 and DGP5 in Table \ref{tab:EOTInt} in the supplementary material. In all cases, the MSREs decrease as the sample size increases. Additionally, the MSREs are smaller for larger values of \(\lambda\), and decrease as \(k\) increases. 
The EoTM-1 performs better in most cases. Notably, when \(\lambda = 1\), the performance of QFM and QRIFE worsens significantly. Two reasons contribute to this poor performance. First, the conditions under which QFM and QRIFE operate are violated when the tail quantile level approaches 0. A critical assumption is that the probability density of \(u_{i,t}\) is bounded around the quantile level where QFM and QRIFE are applied, which is no longer valid as the tail quantile level nears 0 when $\lambda = 1$. Second, since QFM is equivalent to FTVM without the bounded constraint in the optimization problem \eqref{eq:problem2}, QFM may overfit the data. 

   \subsubsection{Mean Squared Relative Error of Extreme Tail Quantiles}



\begin{figure}[htbp]
    \centering
    \includegraphics[width=\linewidth]{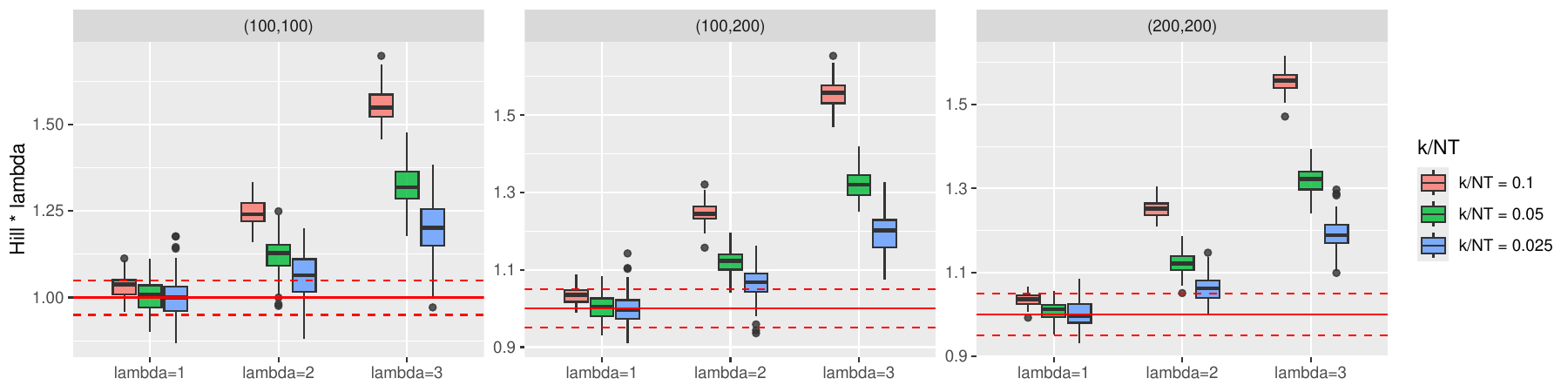}
    \caption{Boxplot of the Hill estimator for different $\lambda$, $k/NT$, and sample size $(N,T)$. We adjust the Hill estimator by multiplying $\lambda$ so that the asymptotic distribution of $\hat{\gamma}_{adj} \lambda$ is a standard normal distribution as $N, T \to\infty$. The solid line represents 1, and the dashed line represents $1.05$ and $0.95$. The data is generated from DGP4.}
    \label{fig:Hill}
\end{figure}

Tables \ref{tab:DGP4EVT} and \ref{tab:DGP5EVT} in supplementary material report the MSREs for DGP4 and DGP5, respectively. For each \(\lambda\), the MSREs initially decrease and then increase as \(k\) grows. This behavior likely results from a trade-off between the increasing MSREs and the decreasing bias of the Hill estimator.  The results in Figure \ref{fig:Hill} show that as \(\lambda\) increases, the bias of the Hill estimator becomes larger. However, reducing \(k\) alleviates this bias. Notably, when \(\lambda = 3\), the bias of the Hill estimator becomes significant, leading to poor performance of the extreme tail quantile estimators when extrapolating from \(k = 0.1NT\). In practice, we recommend using the Hill plot to select an appropriate \(k\).

\begin{figure}[t]
    \centering
    \includegraphics[width=.9\linewidth]{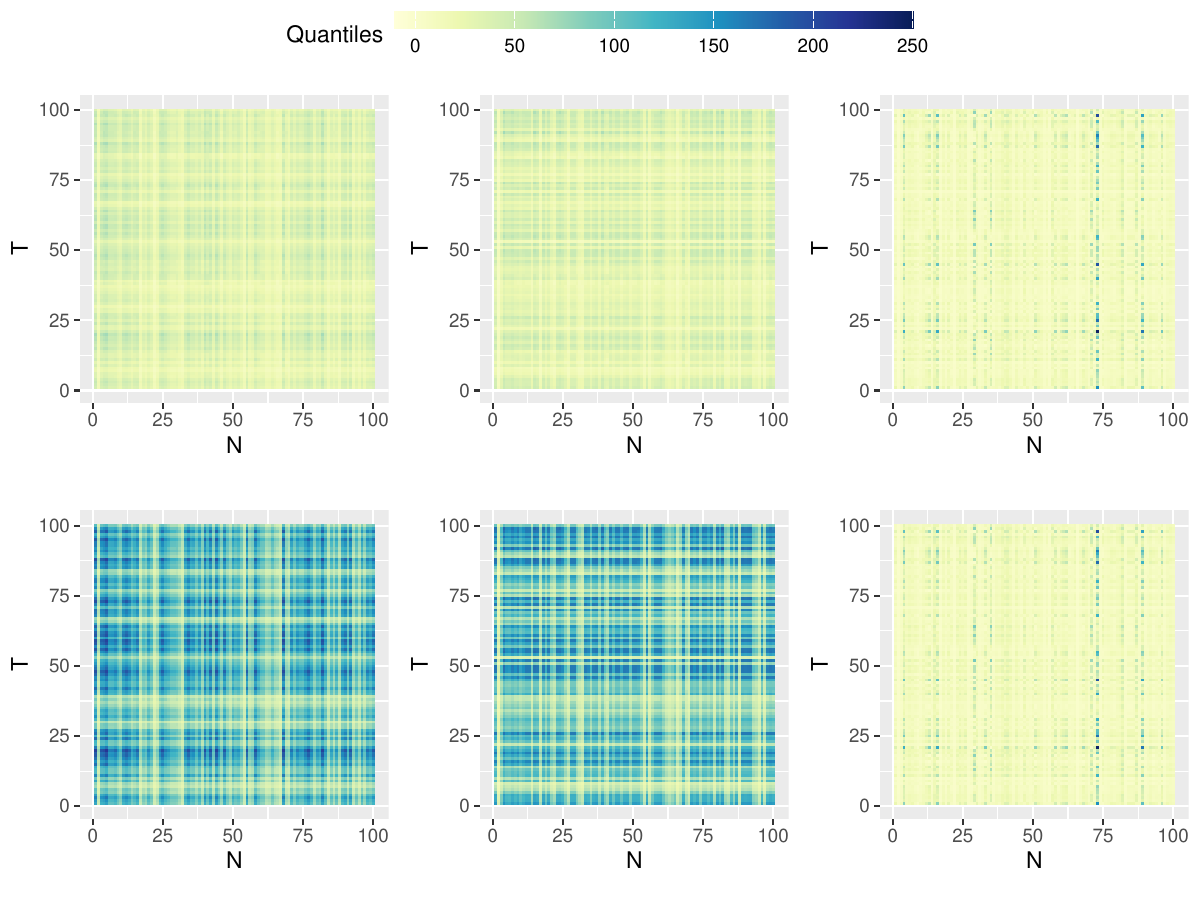}
    \caption{
 For $Y_{i,t}$ generated from DGP4, we plot the 
 quantiles of $Y_{i,t}$ at $p_{N,T} = 0.001$(top-left) and
    $p_{N,T} = 0.0001$(bottom-left), estimated quantiles by EoTM at $p_{N,T} = 0.001$(top-middle) and
    $p_{N,T} = 0.0001$(bottom-middle), and estimated quantiles by QFM at $p_{N,T} = 0.001$(top-right) and $p_{N,T} = 0.0001$(bottom-right).
 }
    \label{fig:UestQFM}
\end{figure}

An interesting observation is that the MSREs of QFM and QRIFE improve at extreme tail quantile levels compared to intermediate tail quantile levels. Evidence from the heatmap in Figure \ref{fig:UestQFM} indicates that QFM estimates similar quantiles at \(p_{N,T} = 0.001\) and \(p_{N,T} = 0.0001\). This improvement can be partially explained by the divergence of \(U_{i,t}(1/p_{N, T})\) to infinity as \(N, T \to \infty\). As the denominator \(U_{i,t}(1/p_{N,T})\) in \(\operatorname{MSRE}_{p_{N,T}}^{\text{EoTM}}\) becomes larger, the value of \(\operatorname{MSRE}_{p_{N,T}}^{\text{EoTM}}\) decreases.
     
\section{Conclusion}\label{sec-conc}

In this article, we introduced the FTVM as a novel framework for modeling and estimating intermediate and extreme tail quantiles in high dimensional data. We also introduce the FTVM-EoT approach to combine the FTVM with other statistical models to connect the relationship between central, intermediate, and extreme quantiles.
To address the challenges of model selection and validation, we developed a hypothesis testing procedure based on the KS statistic and introduced an information criterion for determining the optimal number of factors.  We establish the asymptotic properties of the factors, loadings, and the intermediate and extreme tail quantiles for both models. We also provide the asymptotic properties of the model validation and model selection method.

We conduct several simulation experiments to evaluate the performance of the FTVM and the FTVM-EoT approach under various DGPs. The results demonstrate the robustness and accuracy of the proposed methods in estimating intermediate and extreme tail quantiles, particularly in scenarios involving heavy-tailed distributions.

\section{Disclosure statement}\label{disclosure-statement}

The authors declare no conflicts of interest.



\phantomsection\label{supplementary-material}
\bigskip

\begin{center}

{\large\bf SUPPLEMENTARY MATERIAL}

\end{center}

\begin{description}
\item[Title:] {\bf Supplementary Material for ``Factorized Tail Volatility Model: Augmenting Excess-over-Threshold Method for High-Dimensional Heavy-Tailed Data''} \newline
This document contains the proofs for Theorems \ref{thm:asymrate} and \ref{thm:selection}. It also includes additional numerical results, Tables \ref{tab:EOTInt}, \ref{tab:DGP4EVT}, \ref{tab:DGP5EVT}, and Algorithm \ref{alg:TQFM2} for solving \eqref{eq:problem2}. (pdf)
\end{description}



  \bibliography{Bibliography-MM-MC.bib}

\end{document}